# Free form three dimensional integrated circuits and wearables on a thread using organic eutectogel gated electrochemical transistors

Rachel E. Owyeung[1,2,3], Wenxin Zeng[2,3], Matthew J. Panzer[1], Sameer Sonkusale[2,3]†

[1]Department of Chemical and Biological Engineering, Tufts University, 4 Colby Street, Medford massachusetts 02155, United States

[2]Department of Electrical and Computer Engineering, Tufts University, 161 College Ave, Medford massachusetts 02155, United States

[3]Nano Lab, Advanced Technology Laboratory, Tufts University, 200 Boston Ave. Suite 2600, Medford Massachusetts 02155, United States

†Corresponding author **Email:** sameer@ece.tufts.edu

**Abstract**

Existing flexible electronics are planar, nonbreathable, and easily damaged. In this work, authors introduce a new paradigm of free-form three-dimensional integrated circuits assembled on a single textile thread for ultimate flexibility. Transistors are assembled as daisy chain on a thread and utilize deep eutectic solvent gels, or eutectogels as nonvolatile, breathable, and repairable dielectric for gating organic electrochemical transistor. The eutectogels provide stable transistor performance without the need for encapsulation, are air permeable to the skin, and make the Organic Eutectogel Gated Electrochemical Transistor (OEGET) a promising candidate for wearable electronics. Complex analog integrated circuits, such as single stage and multi-stage amplifiers can be realized on a single thread which can bend out of the plane of stretching for true three-dimensional flexibility. An all-thread based wearable that uses a thread-based strain sensor and thread-based integrated circuit for eye motion and respiration monitoring is also presented.

**Teaser**

A complex three-dimensional integrated circuit on single thread achieves ultimate flexibility and utilizes non-volatile breathable gels for transistor action.

## INTRODUCTION

The flexibility and performance of current bioelectronic interfaces is limited primarily due to the electronic circuits and interconnects used for sensor readout and computing. To increase flexibility, emerging wearable devices use thin-film inorganic transistors (e.g. thin silicon) and thin dielectrics with serpentine interconnects on ultrathin polymeric substrates.(*1, 2*) Others utilize transistors made with organic semiconductors, which are naturally soft and flexible and closely resemble the mechanical properties of the human skin.(*3–5*) Despite the selection of materials and processing to improve the flexibility of transistors and integrated circuits, a mismatch still exists between them and the underlying skin or tissue. This is because existing bioelectronics are limited to two-dimensional planar configurations, which restricts out-of-plane motion and makes it challenging to adapt to the complex, three-dimensional contours of the human body. Additionally, the use of materials that are not air-permeable (e.g., silicon) prevents natural skin breathing. Innovations on multiple fronts are needed to implement truly flexible and breathable bioelectronics.

The *first* innovation addresses the issue of form factor to achieve better interface with any complex 3D body shape. Our solution is a one-dimensional thread or fiber to realize an entire integrated bioelectronic circuit, giving it arbitrary three-dimensional flexibility and stretchability. Compare this to an existing flexible electronics platform that can only flex or stretch in two dimensions as they are bound to a planar substrate. The result is a more resilient and durable interface even under repeated body motion. Figure 1 (a) compares the proposed thread-based bioelectronics with other possible candidates for wearable bioelectronics. There have been promising realizations of transistors on fibers and textiles(*6*), however they are either limited to single transistor demonstrations(*6–11*), or woven into textiles for planar implementations(*6, 11–13*) that are unable to achieve arbitrary 3D flexibility. Prior thread-based transistor realizations utilize photolithography, vapor deposition and/or high temperature cleanroom processes that are not compatible with low temperature polymeric or natural materials such as threads. Moreover, none of the prior implementations have explored free-form 3D realizations of thread-based integrated circuits. The *second* innovation addresses the issue of scalable manufacturing of transistors, sensors and circuits suitable for the curvilinear and/or fibrous nature of the thread, and doing so without cleanroom processing and instead relying on a combination of 3D stencil-based patterning, electroless gold plating and drop-casting. The new method also allows for high throughput realization of other circuit components such as interconnects and resistors in addition to transistors along a single thread for a truly flexible 3D integrated circuit.

The *third* innovation pertains to the choice of materials and device configuration to realize transistor on thread to achieve both superior electronic performance (e.g., high transconductance), and flexibility, breathability, and long-term reliability. This is satisfied by utilizing deep eutectic solvent gels or eutectogels as nonvolatile, breathable, and repairable dielectric for the transistor, but also as breathable anchors for on-skin attachment. This allows the thread-based integrated circuit to flex and stretch both in and out of plane for three-dimensional flexibility as shown in Fig. 1(b) and (c). The eutectogel provides a nonvolatile electrolyte that allows for long-term device usage without the need for additional encapsulation, while the gel scaffold provides mechanical stability, eliminating the requirement for a separate substrate to support electronic devices. Compared to hydrogels, which are prone to dehydration, the nonvolatile nature of the eutectogel ensures stability in ambient conditions for several days or weeks. Additionally, experiments suggest excellent air and water vapor permeability, allowing the eutectogel to remain on the body without hindering underlying skin breathability, which is a crucial metric for any long-term wearable device. Combining thread-based transistors and the resulting thread-based integrated circuit with potential thread-based sensors, can help realize a first of its kind all-thread-based wearable, which can conveniently curve and flex around complex body parts such as around the earlobe as shown in Figure 1(d) or around a finger as shown in Fig. 1(e) and (j). The thread-based free-form integrated circuit are easily anchored on skin via eutectogels as shown in Fig. 1(k) and (l). Fig. 1(f) and (g) demonstrate how the geometrical freedom endowed by the thread could lead to coiled integrated circuits that can also stretch or extend over 16-fold. Further, with simple application of heat, the eutectogels can be fused back together to repair a catastrophically broken device (Fig. 1(i-k)) *vide infra*.

The *fourth* innovation lies in the nature of the transistor itself. It is based on a reimagined design of the popular organic electrochemical transistors (OECTs) by employing eutectogels as an electrolyte gel for transistor gating. OECT are a popular choice for bioelectronics due to their biofriendly composition, ease of processing, and high transconductance.(*14, 15*) It utilizes doped conductive polymer such as poly(3,4-ethylenedioxythiophene) polystyrene sulfonate (PEDOT:PSS), where $PEDOT^{+}$ is volumetrically de-doped by the ions in the gate electrolyte, which is typically an aqueous solution of sodium chloride or similarly small salt, as a function of gate potential. While aqueous-based OECTs are highly effective for sensing in biological environments, their low volatility hinders their use in wearable bioelectronics. Additionally, without proper encapsulation, these devices are prone to leakage when subjected to deformation. Considering these challenges, researchers are exploring alternative electrolytes to aqueous solutions. One option is ionic liquid or ionic gel (ionogel) electrolytes, which exhibit non-volatility due to their high ion density. (*16–18*) However, ionic liquids are often toxic and costly, and require the addition of smaller ionic species such as sodium chloride for transistor action. (*18*) We choose deep eutectic solvents instead, which share many similarities with ionic liquids, including low volatility, adequate ionic conductivity, and chemical tunability. (*19, 20*) Unlike ionic liquids, deep eutectic solvents are typically composed of biofriendly and cost-effective components, making them a viable option for on-skin wearable bioelectronics. (*20*) The most used deep eutectic solvent consists of a hydrogen bond acceptor (HBA) quaternary ammonium salt and a hydrogen bond donor (HBD) such as glycerol, ethylene glycol, or urea. (*19, 21–23*) Interestingly, many of these hydrogen bond donors are also used in PEDOT: PSS formulations to enhance the conductivity in OECTs. (*24, 25*) The result is an Organic Eutectogel Gated Electrochemical Transistor or OEGET, as shown in Fig. 1(h) and (i). OEGETs also show some highly unique and desirable properties. For example, all-around electrostatic gating via gel electrolyte and long-range polarizability (*26*) allows for limitless 3D device geometries. Another example is in Supplementary Information (SI) in Fig. S1 which shows a transistor as a mobius strip. Furthermore, with simple application of heat, the eutectogels can fuse back together if broken with no change in electronic or mechanical performance. This makes OEGET repairable at its mechanically weakest section where eutectogels touch the skin (as anchors), while also serving as the dielectric for all-around transistor gating.

For the *fifth innovation,* we utilize OEGETs high transconductance for the realization of analog circuits, namely amplifiers. Current demonstrations using organic transistors have so far been limited to small-scale digital integrated circuits and biosensors.(*6, 10, 12, 27–30*) There have only been a limited number of implementations of amplifiers (*31, 32*) and complex analog integrated circuits.(*33*). As a proof of concept, we realize multi-stage analog integrated circuits based on OEGETs on a single thread. Moreoover, for the first time, we also demonstrate an all-thread-based wearable that combines thread-based sensors and integrated circuits for eye-motion tracking and respiration monitoring. The resultant all-thread-based integrated circuit and wearable meets our high bar for any bioelectronic device – that of flexibility, softness, compliance, breathability, and adaptability to complex 3D body geometry. (*21*)

## RESULTS

### Fabrication of Eutectogel and Eutectogel Gated Electrochemical Transistor (OEGET)

The eutectogel uses a DES/H2O mixture as a nonvolatile electrolyte with a 1:2:1 molar ratio of choline chloride, ethylene glycol, and water, supported by 20 weight % gelatin. According to a previous report, this eutectogel exhibits superior mechanical and ionic properties compared to similar choline chloride-based DES gelatin gels.(*34*) The tensile modulus was measured to be 26 kPa, with toughness of 68 $kJ/m^3$. The gel is stretchable up to 270%, and has an ionic conductivity of 5.2 mS/cm.(*23*) Additionally, ethylene glycol is already present in many PEDOT:PSS formulations including ours, to induce conformational changes of PEDOT chains, which improves overall conductivity. (*25, 35*) The gelatin scaffold is also essential for facile fabrication. The thermoreversible nature of the physical gelatin crosslinks facilitates encapsulation of the active channel following drop-casting of the gel precursor solution heated above 50℃ on it. Alternately, simply sewing the active channel thread through a pre-molded eutectogel can also be performed. Both approaches provide a diversity of methods to realize highly unique transistor geometries and endow the bulk of the device with on-demand repairability with the application of heat (*vide infra*).

The OEGET functions as a p-type depletion mode device, where it remains ON until a positive gate bias is applied, driving larger cations (choline+) from the eutectogel into the PEDOT:PSS active channel coated on the thread (Fig. 1(i)). The choline+ interacts with PSS-, reducing PEDOT+ to neutral PEDOT. This dedoping effectively decreases the conductivity of the channel, ultimately turning the device OFF at high enough bias. Unlike previous demonstrations of nonaqueous-gated OECTs (13), no additional ions are added, indicating that the bulky cation, choline, is sufficient to provide transistor action. This is the first report to demonstrate choline chloride operation of PEDOT:PSS OECTs, suggesting promise for driving OECTs with larger ions, such as for sensing larger ions or employing other nonvolatile electrolytes.

**Scalable fabrication of OEGET based integrated circuits on thread**

The active channel is formed on threads via a cleanroom-free, three-dimensional stencil-based pattering(*36*) as illustrated schematically in Fig. 2. First, Ecoflex stencil masks are fabricated via doctor blading (Fig. 2(a)). Then, monofilament polycaprolactone (PCL) threads (average diameter = 0.19±0.05 mm) are sewn through a pre-stretched Ecoflex sheet (Fig. 2(c)), such that upon release of the strain applied to the Ecoflex sheet, a tight seal is created around the thread (Fig. 2(d)). The Ecoflex mask serves as a 3D shadow mask for deposition of a conductive coating. The threads embedded in the Ecoflex 3D shadow mask are first surface treated via air plasma (Fig. 2(e)), then silanized to attach amine-rich groups to the surface of the threads (see methods), as shown in Fig. 2(f). After, the thread/mask ensemble is immersed in a solution of negatively charged gold nanoparticles to act as a seed layer for gold deposition (Fig. 2(g)). The nanoparticles promote gold reduction on the threads, where they are seeded.(*37*) Lastly, the 3D-mask/threads ensemble is immersed in a gold salt solution with a reducing agent to electrolessly deposit gold all around the PCL threads (Fig. 2(h)). Following gold deposition, the resulting thread is pulled out of the mask. Tens to hundreds of active channel gaps can be made in a single batch (SI Fig. S2 (a)-(b)) The thickness of the mask correlates to the length of the resulting active channel (SI Fig. S3). Versus previous implementations of active channel gaps fabricated on threads that rely on application of conductive inks or sputtering,(*36, 38*) this approach is cleanroom free, high throughput, and produces uniform, conductive films. Thicker films of gold can be achieved by replenishing the gold salt and reducing agent. Through experimentation, we found that replacing this gold salt solution three times during the deposition process resulted in highly conductive gold PCL threads (see SI Fig. S4), 1.9 ± 0.1 Ω/cm. The average gap, L, between S/D is 0.59±.07 mm when an Ecoflex mask thickness of 0.5 mm is used. The width, W, is defined as the circumference of the PCL thread (πD), where D is the thread diameter, resulting in a W/L ratio of 1.02 ± 0.27 (N=7). The PCL threads can also be sewn as a single, continuous segment to create one thread with multiple gaps (SI Fig. S2(c-d)). These gaps can be subsequently filled with PEDOT:PSS (Fig. 2(i)) and encapsulated with eutectogel through drop casting of gel precursor solution (or sewing through a pre-formed eutectogel). Gating with a thread or fabric electrode realizes an OEGET, and without gating it realizes a resistor.

**Performance of fabricated OEGETs**

Fig. 3(a-b) illustrates the transfer and output characteristics, respectively, of a champion OEGET for an effective width/length ratio (W/L) of 1.2. A carbon fabric (or another conductive thread) is used to apply a gate potential. Some hysteresis, though minor, is observed, likely due to sluggish ion motion, as the hysteresis increases proportionally with sweep rate (see SI Fig. S5). The specific capacitance of the eutectogel as a function of frequency can be found in SI Fig. S6 along with transient current characteristics in SI Fig. S7 and S8. Within a 1V operation window, the ON/OFF ratio is $10^3$ for the champion device, and $10^2$ on average (N = 16). The electrical symbol of the OEGET, and its corresponding small signal model are shown in Fig. 3(c). Here, $r_0$ is the small signal output impedance of OEGET around the biasing point and $r_L$ represents any gate leakage. For planar realizations of OECT or OEGET, a transistor in saturation mode can be described by a simplified MOSFET equation accounting for volumetric capacitance:

$$I_{DS} = \frac{Wd}{2L}\mu C^*(V_G - V_T)^2 \quad (1)$$

Where μ is the charge carrier mobility, C* is the volumetric capacitance, $V_T$ is the threshold voltage, and d is the channel thickness. For 3D realizations such as ours, finding the channel thickness can be problematic. Instead, the volume of the active channel material, $\vartheta$, can be used for a more generalized representation of this equation ($\vartheta = WdL$):

$$I_{DS} = \frac{\vartheta}{2L^2}\mu C^*(V_G - V_T)^2 \quad (2)$$

Equation 2 indicates the transistor action is primarily dependent on the volume of active channel material and gap length. The volume of the active channel material when dried was 0.05 $mm^3$, thus our geometrical equivalent to Wd/L is 140. The small signal output resistance around the biasing point, $r_o$, is 111 kΩ. Maximum transconductance, $g_m = \partial I_D/\partial V_G$, for the champion device reached 16 mS and was 9±5 mS on average (N = 16). A 100 mM sodium chloride-gated PEDOT:PSS showed $g_m$ = 2.7 mS for a similar W/L ratio and PEDOT:PSS thickness of 400 nm.(*39*) Since $g_m$ can be increased simply by increasing PEDOT:PSS thickness, it is common to report $g_m$ with respect to the active channel geometry (for planar devices = Wd/L). Again, for a geometry immune equation 2, we can report $g_m L^2/\vartheta$, which is 0.11 mS/μm for our champion device. A helpful figure of merit when designing organic electrochemical transistors for circuits is the transconductance per drain current, or transconductance efficiency, at an intended biasing point. Our circuits are biased around $V_{GS}$ = 0 V (*vide infra*) to reduce voltages needed, and the average transconductance efficiency was 4.7±1.2 $V^{-1}$ (N = 16).

To showcase the mechanical flexibility of the Organic Electrochemical Transistors (OEGETs), we conducted transfer curve measurements on a single OEGET device at various bending radii, as illustrated in Fig. 3(d). We observed a minor reduction in the maximum current, indicating a slight change in the transistor's performance. Furthermore, the transconductance efficiency at VGS = 0 V was measured as a function of the bending radius (Fig. 3(e)), revealing a slight alteration in transistor performance at lower bending radii, which persisted even up to a bending radius of 6 mm.

To realize integrated circuitsm with multiple transistors, it is necessary to fabricate transistors with various aspect ratios determined by the transistor's effective width, channel length, and the amount of PEDOT:PSS, i.e., $\vartheta/L^2$ or Wd/L. For creating a few discrete transistors, we can vary the thickness of the Ecoflex stencil to obtain transistors with different channel lengths, or use threads of different widths. However, this approach may not be feasible for realizing integrated circuits where transistors of varying aspect ratios need to be fabricated on a single thread. In such cases, we can utilize a standard approach employed in silicon IC design, using series and parallel connections of unit transistors, which is typically a transistor with smallest dimension. For planar devices, when N unit transistors, each with an effective width of $W_u$ and length of $L_u$ are connected in series with shared gate dielectric and a single gate input, we can approximate his to be an effective transistor of width $W_u$ and length $NL_u$. On the other hand, when connected in parallel with all drains shorted together and source connections shorted together, with shared dielectric and a single gate input, twe can approximate this to be a transistor with effective width $NW_u$ and length $L_u$.

For a proof-of-concept demonstration, we fabricated two OEGETs named T1 and T2, and their transfer curves are shown in Fig. 4(a). When connected in series with shared eutectogel and a single gate input, the resulting transfer curve has a reduced current magnitude (Fig. 4(b)). The estimated $I_{DS}$ (series) at $V_{GS}$ = -0.5 V differs from the experimental results by 10%. Conversely, when T1 and T2 are connected in parallel with shared dielectric and a single gate input, the resulting transfer curve has increased current magnitude (Fig. 4(c)). The estimated $I_{DS}$ (parallel) at $V_{GS}$ = -0.5 V differs from the experimental results by only 7%. The difference between theoretical and experimental values may be due to a small gate leakage current, resulting in a small IR voltage drop across the gate connections. This simple proof-of-concept can be extended to any combination of series and parallel realizations for realizing any arbitrary effective width-to-length ratios to aid in circuit design. Fig. 4(d) shows an actual device of a 5 series connected transistors sharing the same gel dielectric and gate, for an effective aspect ratio of $W_u/5L_u$.

**Repairability and Breathability of Eutectogel and OEGETs**

The eutectogels exhibit mechanical flexibility due to the dynamic interactions between gelatin triple helices and nonhelical interchain hydrogen bonding.(*23*) These interactions are thermoreversible, making the eutectogels easily repairable by the simple application of heat. The healing feature is particularly important for wearable bioelectronic applications, where devices are often subjected to frequent mechanical contact, motion, and strain. Even without heat, the capacitive functionality of the gel in OEGETs can be restored simply by physically reconnecting the broken pieces of the gel. This concept is demonstrated in Fig. 5(a), which shows the capacitance of a eutectogel during a catastrophic breaking (i.e., a cut in half) indicated by event “1.” At event “2,” the two ends of the gel are placed in physical contact with each other, immediately restoring most of the capacitance. At event “3,” the broken interface is gently heated using an external heat gun to initiate the mechanical repair. Photographs of the corresponding eutectogels during the three events are shown in Fig. 5(a) insets. The minor difference in capacitance after event “3” versus the initial capacitance is due to the heat gun causing a change in the cross-sectional area by melting the bulk gel during the experiment.

To study the mechanical repairability further, we measured the series resistance of the eutectogels while stretching the gel until it breaks (Fig. 5(b)). The gel can be stretched to around 240% before breaking. The gel is then repaired by external application of heat (either via a heat source directly or by applying a small amount of preheated gel solution) to the broken interface, and the two ends of the gel are pressed back together, where it is left to cool (for 3 hours). Once fully cooled, we perform the same test to observe how the gel behaves mechanically. The repaired gel breaks at 245% strain. The similarity in the strain-to-break percentage indicates that the gel network has been successfully repaired. Note that the healing process was performed at room temperature and can be accelerated by chilling the gel. Furthermore, Fig. 5(e-g) showcases the catastrophic breaking and successful repairing of a “5-series connected transistor” shown initially in Fig. 4(d).

One also needs to monitor the longitudinal performance of a chronic wearable bioelectronic platform over multiple days. One major advantage of our use of eutectogel is its low volatility, due to attractive interactions between the HBA and the HBDs. As a result, OEGETs will not evaporate like aqueous systems when left exposed to ambient conditions. To further examine the benefits of low volatility electrolytes, we created a hydrogel equivalent of our eutectogels for comparison. This hydrogel had the same concentration of choline chloride and gelatin scaffold as the eutectogel (4 M and 20 wt.%, respectively). The photographs of the two equivalent gels (Fig. 5(h)) show that the hydrogel has a significant volume reduction compared to the eutectogels over time. The hydrogel's volume decreased by 57% after 24 hours and continued to shrink to a volume loss of 75% after 72 hours, whereas there was no detectable volume change for the eutectogels.

We fabricated two separate devices, one with hydrogel gating (Fig. 5(c)) and the other with eutectogel gating (Fig. 5(d)), and recorded transfer and output curves for each over the next few days. Between measurements, the devices were kept at ambient laboratory conditions (22-23℃ and 21-23% RH). From the transfer curves shown in Fig. 5(c-d), it is evident that there is a substantial decline in the device performance after 24 hours for the hydrogel gated device (94% change in $g_m$ within a single day), whereas the eutectogels gated transistor maintains comparable performance over seven days (with the maximum $g_m$ exhibiting only a 12% change after seven days). After a week, the device performance deteriorates; however, the ON current is still higher than the value the hydrogel device achieved after 24 hours ($g_m$ of OEGET after 37 days was 0.28 mS, $g_m$ of hydrogel OECT after 1 day was 0.15 mS).

It is worth noting that the PEDOT:PSS used in all experiments did not have any crosslinker, such as (3-glycidyloxypropyl) trimethoxysilane (GOPS), to enhance the inherent stability of PEDOT:PSS itself. Moreover, the PEDOT:PSS was in contact with the electrolyte continuously for all measurements. Any minor decrease in device performance is likely due to the instability of PEDOT:PSS in ambient conditions, rather than the eutectogel itself.(*25*)

To assess long-term wearable comfort on the skin, another important metric is air/vapor permeability. We conducted water vapor permeability studies, and the results are depicted in Fig. 5(i). We placed known amounts of DI water in vials without a lid (Open (Control)), with a polyethylene terephthalate (PET (Control)) lid, and with a eutectogel lid (~1 mm thick) to assess whether the gel inhibited water vapor from escaping the vial. The total mass of each vial was weighed for 7 days, and the mass change from the initial water weight is plotted in Fig 5(i). As expected, the open vial continuously lost water over the 7 days, and the PET sheet sealed the vial, preventing any measurable amount of water from leaving the container. The eutectogel lid performed similarly to the open container, indicating that it has excellent water vapor permeability. Thus, it will enable the skin to breathe naturally even when covering large areas.(*40*) Moreover, we conducted cell viability studies with the eutectogel to examine its cytotoxicity. We observed a negligible reduction in cell viability for the wells treated with gels up to 10 mg/mL, indicating that the gels are biocompatible when used at a concentration below 10 mg/mL. This value is five times higher than the final concentrations used for work on capacitive ionic skin gel sensors, which only showed concentrations up to 2 mg/mL.(*41*)

**OEGET Integrated circuits on thread**

The high transconductance of OEGETs and the ability to realize multiple transistors on a single thread lends itself to a new paradigm of analog integrated circuits on a single thread. As a first demonstration, an OEGET is implemented into a common source amplifier with a load resistor $R_D$ all implemented on a single thread (Fig. 6(a)). Three active channels are fabricated using the cleanroom free 3D shadow mask approach followed by gold deposition and drop-casting with PEDOT:PSS solution as discussed earlier. Only one channel is further coated with the eutectogel and a gate fabric is added to apply input, thus creating a single OEGET. The gate eutectogel also serves as a breathable anchor for skin attachment. The two remaining PEDOT:PSS channels serve as resistors in series for a combined load resistance for the amplifier. The gain of a single stage common source amplifier shown in Fig. 6(b) schematic is given by $-g_m$ ($R_D$ || $r_0$), where $g_m$ is the small signal transconductance, and $r_0$ is the small signal output impedance of the OEGET around the biasing point (see Fig. 3(c)). To maximize $g_m$ while keeping $I_D$ small, the DC bias at gate, $V_{in}$ (or $V_{GS}$) is chosen around 0 V. The values of $R_D$ and $V_{DD}$ are chosen such that the OEGET operates in the saturation region, yet $V_{DS}$ is not large enough to cause the breakdown of the OEGET which occurs >1.2 V. The resistance of the two PEDOT:PSS resistors in series was 2.19 kΩ total. Here, $V_{DD}$ was kept sufficiently low, at -1.5 V. The theoretical gain for this schematic was calculated to be 3.28 V/V, which matches well with experimental values for gain, at 3.14 V/V (shown in Fig. 6(d)). To increase gain, $V_{DD}$ and $R_D$, or the overall width of the transistor can be increased. This is easily achieved by using parallel connection of multiple active transistor channels, as described earlier. This gain is dependent on frequency as shown in Fig. 6(e). The dominant contributor to limited frequency bandwidth is the large time constant at the gate due to the large electrolytic double layer capacitance and the ionic resistance through the eutectogel. To trade off gain against reasonable supply voltage, $V_{DD}$ was set to -8.8 V, which resulted in a gain of 14 V/V at 500 mHz. Unity gain bandwidth occurs around 15 Hz, with a gain of 20 V/V at the lowest frequency teste of , 50 mHz.

To demonstrate a more complex thread-based integrated circuit using OEGETs, we implement a two-stage cascade amplifier, shown schematically in Fig. 6(c). Note that a single thread can be patterned to realize the entire analog integrated circuit by creating 6 gaps using the 3d shadow mask patterning, and drop-casting PEDOT:PSS on all of them. Eutectogel encapsulation and voltage gating is performed on two of the six channels to realize transistors $T_1$ and $T_2$, while others are not gated to realize resistances $R_{D1}$, $R_3$, $R_4$, $R_{D2}$ in the following sequence $T_1$ – $R_{D1}$ – $R_4$ – $R_3$ – $T_2$ – $R_{D2}$. They are then interconnected for ground and power supply connections as shown. Eutectogels used to realize $T_1$ and $T_2$ also provide natural breathable anchors on the skin without the need for any substrate support. Note that both stages resemble a common source amplifier on their own, and they are capacitively coupled ($C_1$) (externally) to only permit the AC signal to pass through to the second stage. In doing so, we can maintain the optimal biasing point ($V_{GS}$ = 0 V) for the second stage, which is set by a voltage divider through resistors $R_3$ and $R_4$ ($R_3 = R_4 = 110\ k\Omega$). The resulting gain at 500 mHz is 34 V/V, versus 14 V/V from the single stage at the same frequency. This is the first multistage analog amplifier realized using organic electrochemical transistors for high gain.

Thread-based transistors, render a new class of truly flexible, breathable, three-dimensional all-thread based wearable when integrated with a thread-based sensor. Towards an example of an all-thread-based wearable, we combine a thread based OEGET amplifier with a thread-based strain sensor. Strain sensors are made through conductive carbon ink coating of elastic threads further sealed with a thin layer of Ecoflex to protect the carbon coating.(*42, 43*) For a practical demonstration, we employ a thread-based resistive strain sensor (with gauge factor in the range of 1.2~2.1) to monitor body motion. The electrical schematic of the circuit is shown in Fig. 7(a). Here, $V_{DD} = \pm 1.6V$ to keep voltage reasonably low for on-body usage, while $R_D = 3.7\ k\Omega$ and $R_3 = R_4 = 110\ k\Omega$. An off-the-shelf potentiometer (not thread-based) is connected in series with the sensor, $R_S$, which allows us to adjust sensor sensitivity. This ensures the gate of the OEGET is not biased at a point that would exceed the electrochemical stability window of the eutectogel. We first placed the sensor across the subject's temple using eutectogels as on-skin breathable anchors to monitor eye motion due to blinking for example, as shown in Fig. 7(b-c). The entire platform is free-form (without substrate), and can easily wrap or conform to the 3D contour as shown. The muscle movement due to eye blinking is captured by the strain sensor, and amplified by the circuit, as plotted in Fig. 7(b).

An all-thread based wearable combining an OEGET circuit and thread-based strain sensor can also be placed on the subject's diaphragm to monitor breathing, in a manner similar to where eutectogel is the only supporting contact on skin, ensuring maximum breathability, as illustrated in Fig. 7(f). From the data shown in Fig. 7(d), inhale and exhale patterns can be distinguished, as well as different rates of breathing, as shown in Fig. 7(e). These modest demonstrations can easily be extended to monitor different body motions such as head motion(*44*) and biopotential monitoring in future work.

## DISCUSSION

This paper presents a thread-based transistor and integrated circuit fabricated using a cleanroom free approach to realize an integrated circuits that can flex arbitrarily in three dimensions. Apart from the substrate-free thread-like form factor of the circuit platform, there are many other unique contributions.

First this paper presents a unique organic electrochemical transistor (OECT) gated with a nonvolatile deep eutectic solvent gel namely eutectogel as a gate dielectric. This is the first report to demonstrate choline chloride operation of PEDOT:PSS based OECT. This differs from prior work that utilized choline-based electrolytes, which were predominantly electrostatically driven devices utilizing a counter-ion to choline for transistor action(*45, 46*) or were using choline-based ionic liquid as an additive to PEDOT:PSS in a role similar to ethylene glycol(*13*). Also unique and noteworthy is that gels made from DES also endows the resulting transistors with greater long-term stability compared to archetypical aqueous electrolytes since the electrolyte does not evaporate at room temperature conditions. Compared to other longitudinal studies (see SI Table S1), the device presented here is under constant contact with the eutectogel electrolyte and does not have a crosslinker in the PEDOT:PSS solution to improve PEDOT:PSS stability against degradation as is often employed in literature. Still, the device can last over a month in ambient conditions with minor performance degradation at the end of the month. This late term degradation is likely due to PEDOT:PSS exposure to small amounts of water continuously exposed throughout the experiment. For future students where longitudinal device performance is paramount, the PEDOT:PSS solution could be crosslinked(*47*) to impede swelling from ambient moisture, and the PCL thread could be replaced with a another nonbiodegradable polymer or natural thread.

As the eutectogel does not evaporate, there is no need for further encapsulation of the OEGETs, and they are even air-permeable. This characteristic allows for long-term on-body usage as the underlying skin can still breathe. Moreover, in the proposed free-form realization, the eutectogels serve the dual purpose of providing minimalist on-skin anchors minimizing overall skin contact area compared to any other flexible electronics platform. This feature reduces any potential adverse impacts on the skin. Additionally, cytotoxicity studies on eutectogels indicate that the resulting OEGET is biocompatible. To the best of our knowledge, this is the first organic electrochemical transistor capable of direct on-skin placement. While these results are promising, further skin patch tests may be necessary to observe the long-term effects (over months) of the eutectogel when worn on-body.

The eutectogel is held together by noncovalent, thermoreversible gelatin interactions, which allows the device to be repaired by simple application of heat. Capacitive performance, paramount to OECT operation, can be restored by contact of broken gel pieces, and can further be mechanically restored by heating the broken interface and waiting for the gelatin interactions to reform upon cooling. Existing examples of self-healing properties (*48, 49*) have focused only on the active channel material, and are thus inadequate in by themselves. These prior approaches could employ the eutectogels' thermoreversible property to realize a fully repairable transistor.

The gelatin scaffold further provides leak-proof operation of the device, such that these can be molded onto the 3D contours of biological surfaces, for free-form transistors and three-dimensional integrated circuit geometries. This, combined with the thread-based active channel and resistive circuit elements enables realization of transistor shapes not possible (SI Fig. S1). Further, the thread and gel ensemble are flexible, and resulting OEGETs can be bent to a radius of 6 mm before losing substantial device performance.

Moreover OEGET demonstrates excellent long term reliable performance. The OEGETs exhibit a maximum transconductance of 16 mS for the champion device, which is comparable to high performing planar OECTs (see SI Table S1), even with the use of a bulky cation, choline, rather than sodium to de-dope PEDOT. The resulting OEGET-based single stage and multi-stage amplifiers achieve gains of 14 V/V and 34 V/V for 500 mHz, respectively, which are among the highest reported gains for PEDOT:PSS-based amplifiers on flexible, 3D substrates.(*32*) and greatly exceeds reported values when considering amplifier demonstrations on flexible substrates. (*49–51*) An important point to note is that this is achieved without truly optimizing the effective width or length of the transistor. To achieve higher gain, one could add multiple transistors in parallel connection for each stage and/or optimize the load resistance and power supply.

Our results show transistors and circuits have a modest bandwidth performance, which is adequate for monitoring physical activity such as from respiration or eye movement. One can reduce gel thickness to reduce the ionic resistance through the gel and thus the RC time constant at the gate. Moreover, one can utilize parallel connection of transistors to increase the transconductance and thus speed. Utilizing smaller channel length transistors by choosing a thin shadow mask during fabrication, is also beneficial as it increases the transconductance (inversely proportional to $L^2$ for same volume $\vartheta$ of of the active organic semiconductor) and thus the frequency bandwidth. Also, since only p-type depletion mode transistor was demonstrated in this report, it can be extended to high mobility solution processable organic and inorganic semiconducting materials.

The scalability afforded by stencil based patterning and dropcasting makes it possible to realize really complex analog and digital integrated circuits on threads with tens or hundreds of transistors and resistors. Beyond amplifiers, one could implement wireless transceiver circuits or implement advanced instrumentation circuitry using these transistors. The presented thread-based platform offers a unique combination of essential metrics for flexible bioelectronics – softness, compliance, breathability, repairability and adaptability to any 3D geometry. When integrated with thread-based sensors, one can realize an all-thread based wearable that can adapt to any three-dimensional body contour to form reliable, breathable interfaces with human body for long term longitudinal monitoring.

## MATERIALS AND METHODS

### Materials

All materials were used as received without further purification. Hydrogen tetrachloroaurate (III) trihydrate ($HAuCl_4$-$3H_2O$), sodium citrate trihydrate, 3-aminopropyl-triethoxysilane (APTES), hydroxylamine hydrochloride, poly(3,4-ethylenedioxythiophene)-poly(styrenesulfonate) 3.0-4.0% in H2O, dodecylbenzenesulfonic acid (DBSA), gelatin from porcine skin (Type A, gel strength ~175g Bloom), and polycaprolactone (PCL) were purchased from Sigma Aldrich and used without further purification. Ethylene glycol (99%, Acros Organics) and Choline chloride (ChCl) (JT Baker) were purchased from Fisher Scientific. ChCl was stored in a recirculating nitrogen-filled glovebox ($O_2$, $H_2O$ < 0.1 ppm) to control water absorption from the ambient environment. Spectracarb 2225-Type 900 carbon fabric was purchased from Engineered Fibers Technology LLC.

**High throughput fabrication of active transistor threads**

Please see Fig. 2 for detailed schematics for fabrication of thread-based transistors. A 0.5 mm Ecoflex 00-30 sheet is cast and doctor bladed to form a uniform film, and cured at 60°C. PCL threads were drawn from melt to a diameter of 0.19±0.05 mm.

Colloidal gold was made by first heating a 2.5 x $10^{-4}$ M hydrogen tetrachloroaurate (III) trihydrate solution to boil (100 mL of DI). Then, 3 mL of a 1 wt % sodium citrate trihydrate aqueous solution was added, upon which the solution changed from pale yellow to dark purple. The solution was heated and stirred until the solution changed to a deep red color, and then stirred for an additional 30 min. The solution was kept at 4°C until use. The nanoparticles show a strong absorbance peak around 520 nm (surface plasmon band), observed by UV/VIS (Evolution 220 Thermo Fisher) (see SI Fig. S9).

After silanization, the 3D shadow mask/thread ensemble was immersed in the colloidal gold solution and gently agitated on a shake plate for 25 minutes. Then the ensemble was gently rinsed with DI water before adding the electroless plating solution.

The plating solution consisted of 1 mL of a 5 mM hydrogen tetrachloroaurate (III) trihydrate solution and 50 µL of a 40 mM hydroxylamine hydrochloride solution, diluted with DI water to bring the final volume to 5.5 mL. The 3D shadow mask/thread ensemble was again immersed and gently agitated on a shake plate for 20 minutes. The solution was replaced for a total of 3 coatings, each for 20 minutes to increase the thickness of gold coating (see SI Fig. S4)

PEDOT:PSS solution of 6 vol % ethylene glycol and 0.2 vol % DBSA vortexed and 0.5 µL was dropcast onto the source/drain channel gap of the PCL thread and left to air dry. This corresponds to a volume of approximately 0.05 $mm^3$ when dried.

**Eutectogel synthesis**

Eutectogels were prepared as described in a previous report.(*23*) Briefly, the DES/H2O mixture was first formed by mixing ChCl, EG, and deionized $H_2O$ in a 1:2:1 molar ratio and stirred vigorously, covered, at 90°C for 2 h or until a clear, homogeneous solution was formed. Then to form the gel, gelatin was placed in a separate vial with approximately one fifth of the total DES/mixture volume to be added to “bloom” the gelatin for facile dissolution. The DES/mixture was heated to 80°C was added to the bloomed gelatin for a final 20 wt % gelatin solution. This solution was stirred at 80°C for 1 h or until a transparent solution was formed. The solution was then poured into molds and chilled at 4°C in a covered container for at least 24 h. The solution could be reheated and recast as needed, or directly drop casted onto the active channel when heated.

**Electrical Characterization**

All transistor devices were tested using an Agilent 4156A semiconductor analyzer at room temperature and atmospheric pressure in ambient conditions. AC impedance spectra were measured using a VersaSTAT 3 Potentiostat with a built-in frequency response analyzer (Princeton Applied Research, Oak Ridge, TN). Circuit data was captured by an Agilent MSO6104A mixed signal oscilloscope. $V_{DD}$ was supplied by a Rigol DP1308A DC Power Supply, and $V_{in}$ was generated by a Tektronix AFG 3022B arbitrary function generator. To measure transient current characteristics as a function of gate voltage pulses at 1Hz and 5Hz, a custom transimpedance amplifier made using an operational amplifier and resistor was used (Fig. S8).

**Cell Viability Studies**

Fibroblast 3T3 cells were seeded at 10k at 0.2 mL per well in 96-well plates 24 h before the experiment. The next day, cells were washed with a PBS buffer three times and incubated with the gels at a serial concentration of 1-20 mg/mL in cell culture media for 24 h. The cells were then completely washed off with a PBS buffer three times and 10% alamarBlue in cell culture media was added to each well (220 µL) and incubated further at 37°C for 2 h. Cell viability was then determined by measuring the fluorescence intensity at excitation wavelength 570 nm and emission wavelength 590 nm using a Biotek Synergy H4 spectrophotometer.

## Acknowledgments

The authors would like to thank Junfei Xia for optimizing the electroless gold plating approach and Riddha Das for performing cell viability studies. The authors would also like to thank Hojat Rezaei Nejad for some of the conceptual figure illustrations.

**Funding**

United States National Science Foundation DGE-1144591: REO, MJP, SS
United States National Science Foundation 1931978: REO, WZ, SS
United States National Science Foundation 1935555: REO, SS,
United States Department of Defense Congressionally Directed Peer Review Medical Research Program PRMRP W81XWH-21-2-0012: REO, SS
United States Office of Naval Research DURIP N00014-20-1-2188: SS

**Author Contributions:**

Conceptualization of Eutectogel gated transistor: REO, MJP, SS
Conceptualization of thread-based integrated circuits: SS, REO
Conceptualization of thread-based wearable: SS, REO, WZ
Design of Experiments: REO, WZ
Investigation: REO, SS
Visualization: REO, SS, WZ
Supervision: SS, MJP
Writing—original draft: REO, SS
Writing—review & editing: REO, SS, MJP

**Competing Interest Statement:**

REO, MJP and SS have filed a patent application through the Tufts Technology Transfer Office.

**Data and Material Availability:**

All data are available in the main text or the supplementary materials.

**Figures and Tables**

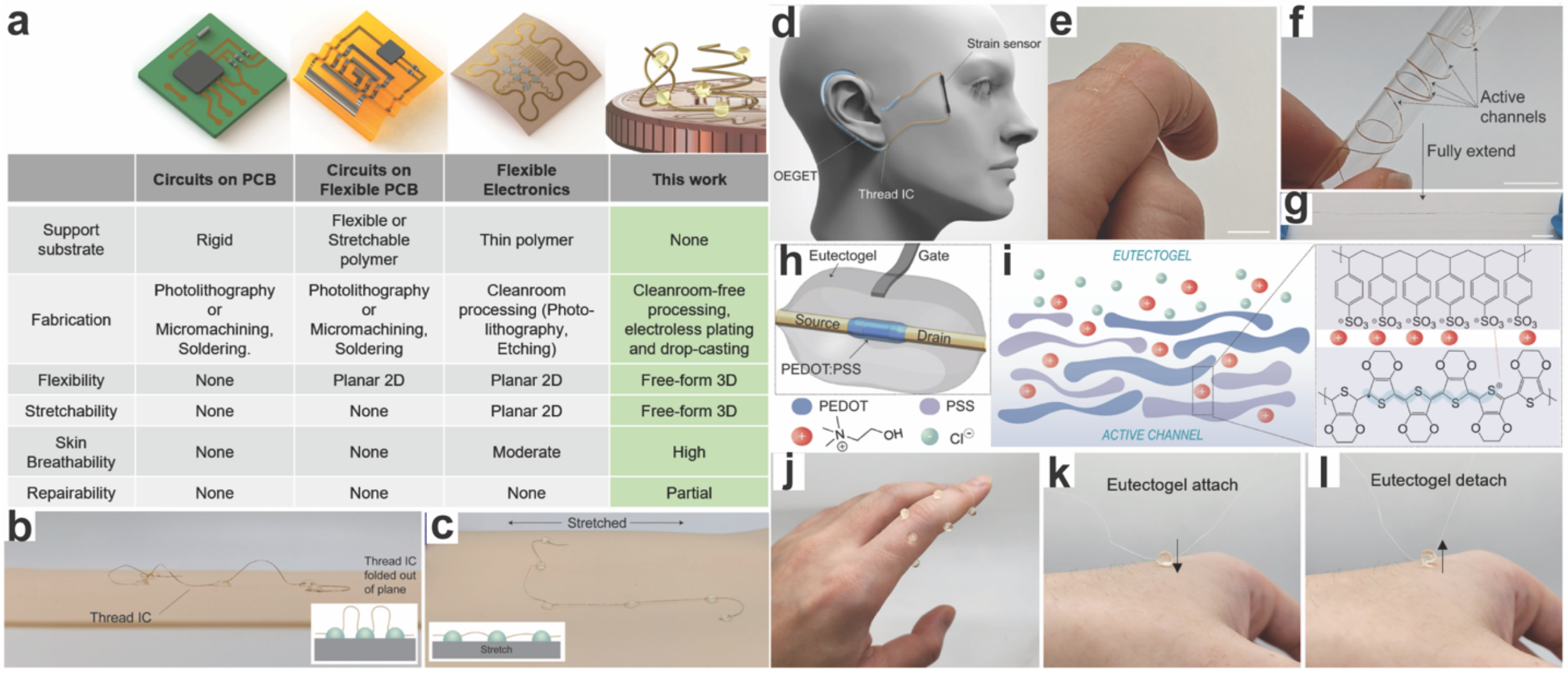

**Figure 1. Conceptual overview and photographs of OEGETs. a.** Comparison of free-form thread-based integrated circuits platform with existing platforms based on printed circuit board and flexible electronics technology. **b.** Demonstration of a single thread integrated circuit (thread IC) comprised of eutectogel gated OECTs (OEGETs) on a skin model. The proposed platform enables the thread IC to be folded out of plane of the human body, such that when the underlying skin is **c.** stretched, the thread IC can lengthen to accommodate the excess strain on the electronics. The eutectogels serve not only as a gate electrolyte but as minimalist breathable anchors for skin attachment as shown. **d.** Conceptual use-case of a free-form thread IC using OEGETs. The single thread OEGET IC is shown to interface with a strain sensing thread in the context of monitoring eye motion as an example application. Free-form realizations allows thread-based OEGET ICs to adapt to any complex contour, such as an ear lobe. **e.** Example OEGET conforming to human skin. **f.** Coiled (8 mm across) and **g.** fully stretched thread (129 mm across) showing 16-fold elongation (stretchability). **h.** Zoomed in view of OEGET cartoon schematic. The OEGET consists of a single thread with gold coated drain and source electrodes, separate by a micro-scale gap where PEDOT:PSS is drop casted. A choline chloride eutectogel provides electrolytic gating of the transistor. **i.** Zoomed in view of PEDOT:PSS active channel and eutectogel interface, showing dedoping of PEDOT:PSS by Choline ions from the eutectogel. **j.** Real life picture of OEGET IC coiled around a finger with eutectogels anchored on skin **k.** Close-up view of eutectogels serving as a minimalist breathable anchor **l.** Simple attachment/detachment of eutectogels anchors All scale bars are 1 cm.

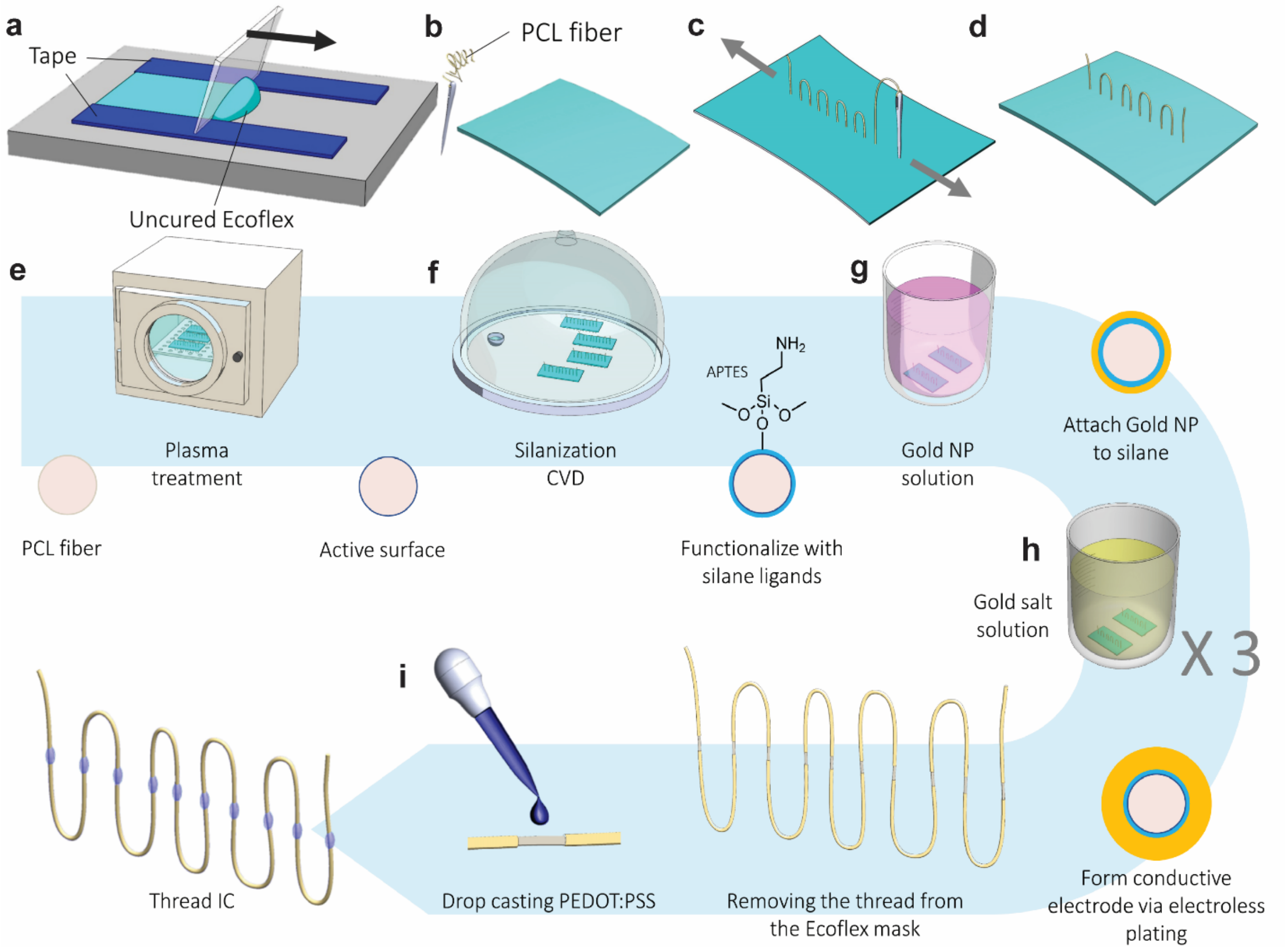

**Figure 2. Cleanroom free fabrication of thread-based integrated circuit based on daisy connected OEGET.** Multiple OEGETs are fabricated using a 3D stencil based patterning of a single thread, namely polycaprolactone (PCL) in this case, followed by electroless deposition of source/drain gold electrodes and drop casting of PEDOT:PSS in a series of following steps **a.** Ecoflex stencil masks are fabricated via doctor blading to desired transistor gap thickness. **b.** Resulting Ecoflex mask and PCL fibers as threads. **c.** Threads are sutured into the pre-stretched Ecoflex mask**. d.** Strain on the Ecoflex mask is released to seal the openings created by suturing. **e.** The thread/mask ensemble is plasma treated, **f.** silanized, and **g.** placed in a bath of gold nanoparticles (NPs) to functionalize any exposed surface with gold NPs. **h.** The thread/mask ensemble is placed in a gold salt solution for electroless deposition of gold all around the exposed surfaces. This forms the drain and source terminals in case of transistors, or contact electrodes in case of resistors **i.** When the thread is pulled out of the mask, the active channel is revealed, which is then coated with PEDOT:PSS via drop casting. The resulting thread is then sewn into pre-formed eutectogels, or have eutectogel precursor solution drop casted on the channels to make OEGETs (not shown).

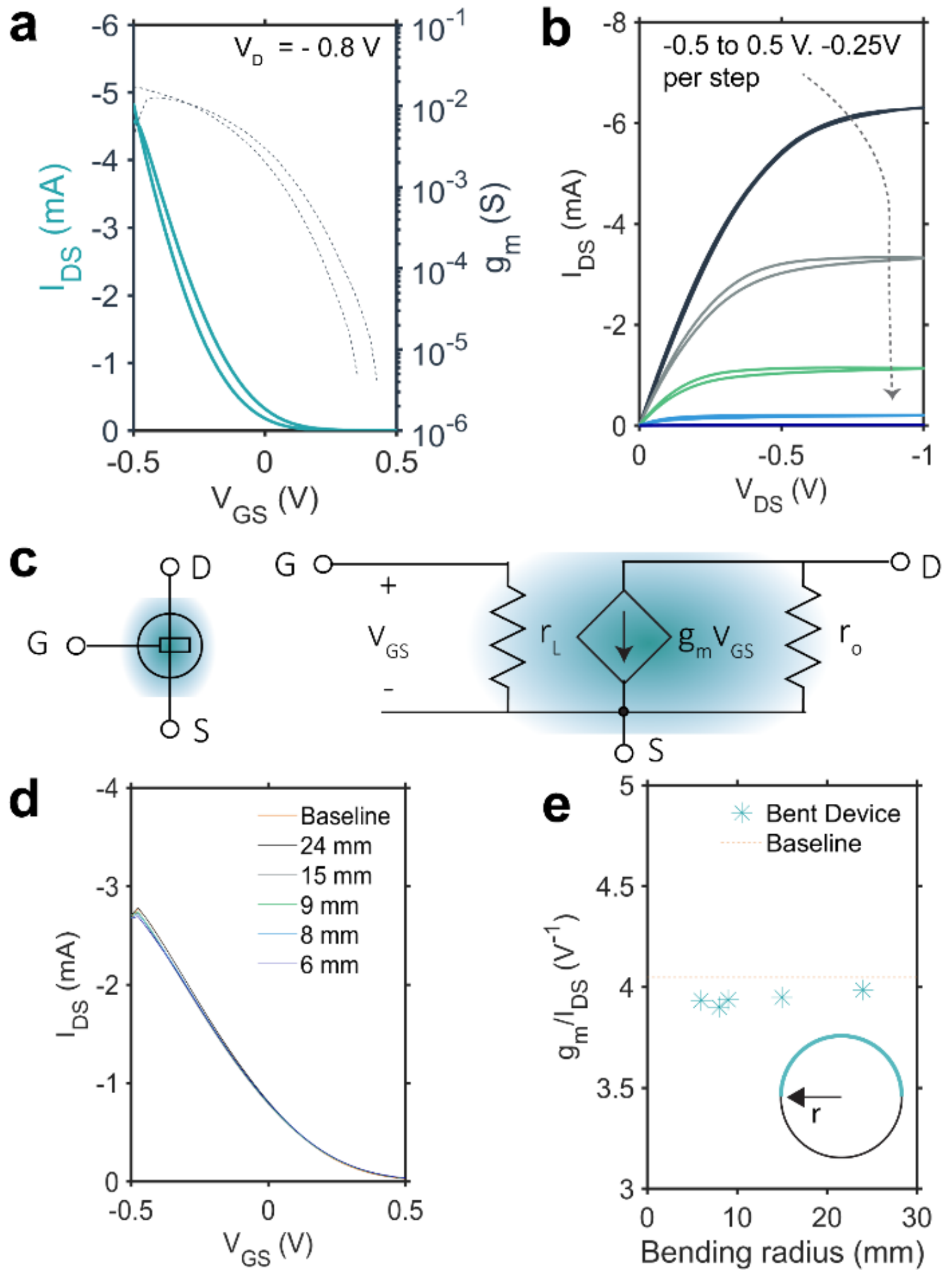

**Figure 3. OEGET characteristics. a.** Transfer and **b.** output characteristic curves of the champion OEGET. **c.** Electrical schematic and corresponding small signal model of the OEGET, where $v_{gs}$ is the small signal voltage change. **d.** Transfer curves of an OEGET at different bending radii. **e.** Transconductance efficiency as a function of bending radius for an OEGET.

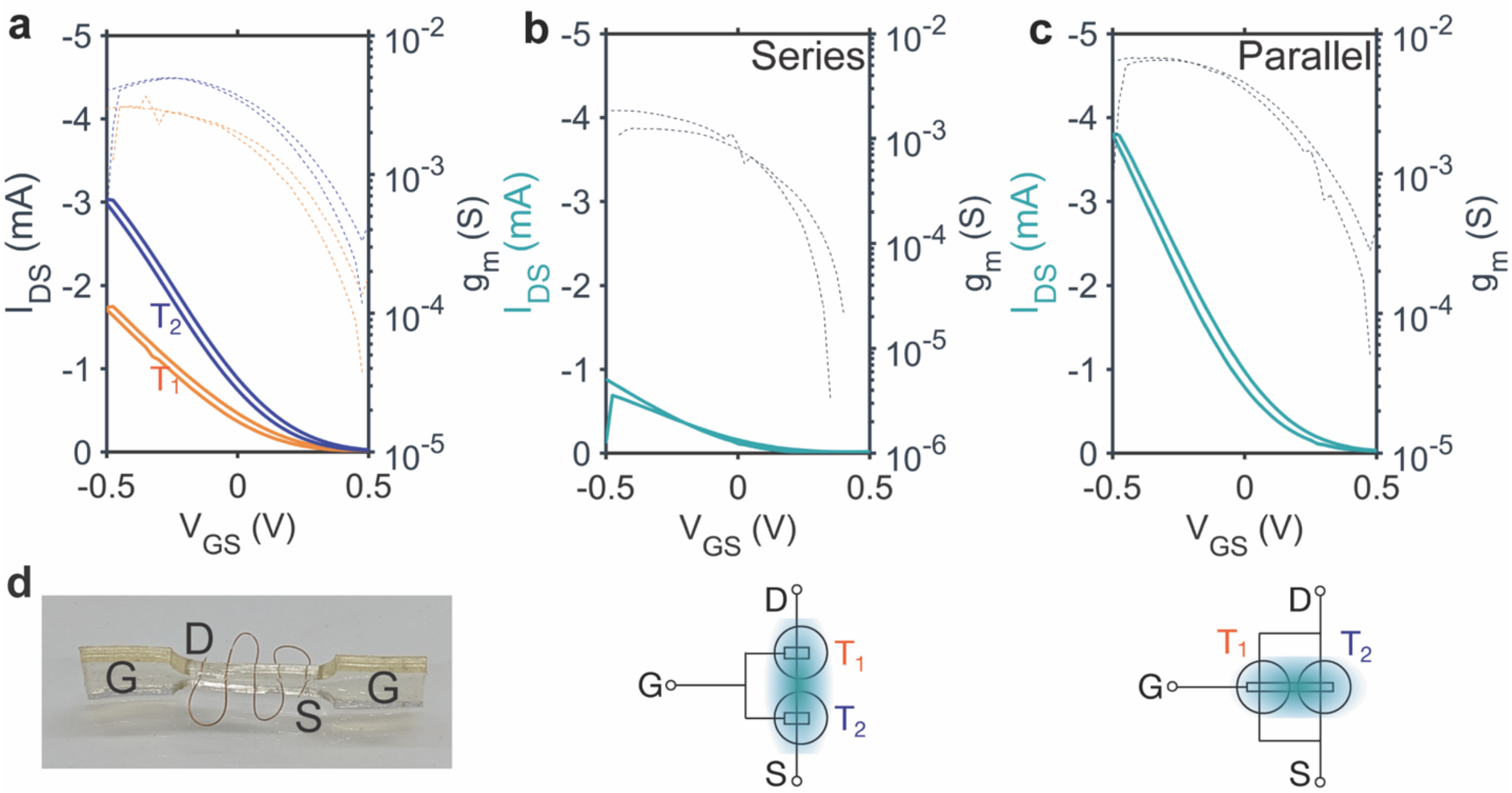

**Figure 4. Series and Parallel demonstration. a.** Transfer curves of individual transistors, $T_1$ and $T_2$. **b**. Resulting transistor curve from $T_1$ and $T_2$ connected in series, sharing the same gate with corresponding electrical schematic (gel overlap shaded in teal). **c**. Resulting transfer curve from $T_1$ and $T_2$ connected in parallel, sharing the same gate with corresponding electrical schematic (gel overlap shaded in teal). **d.** Example realization of a combined 5-unit transistor series-connected OEGET

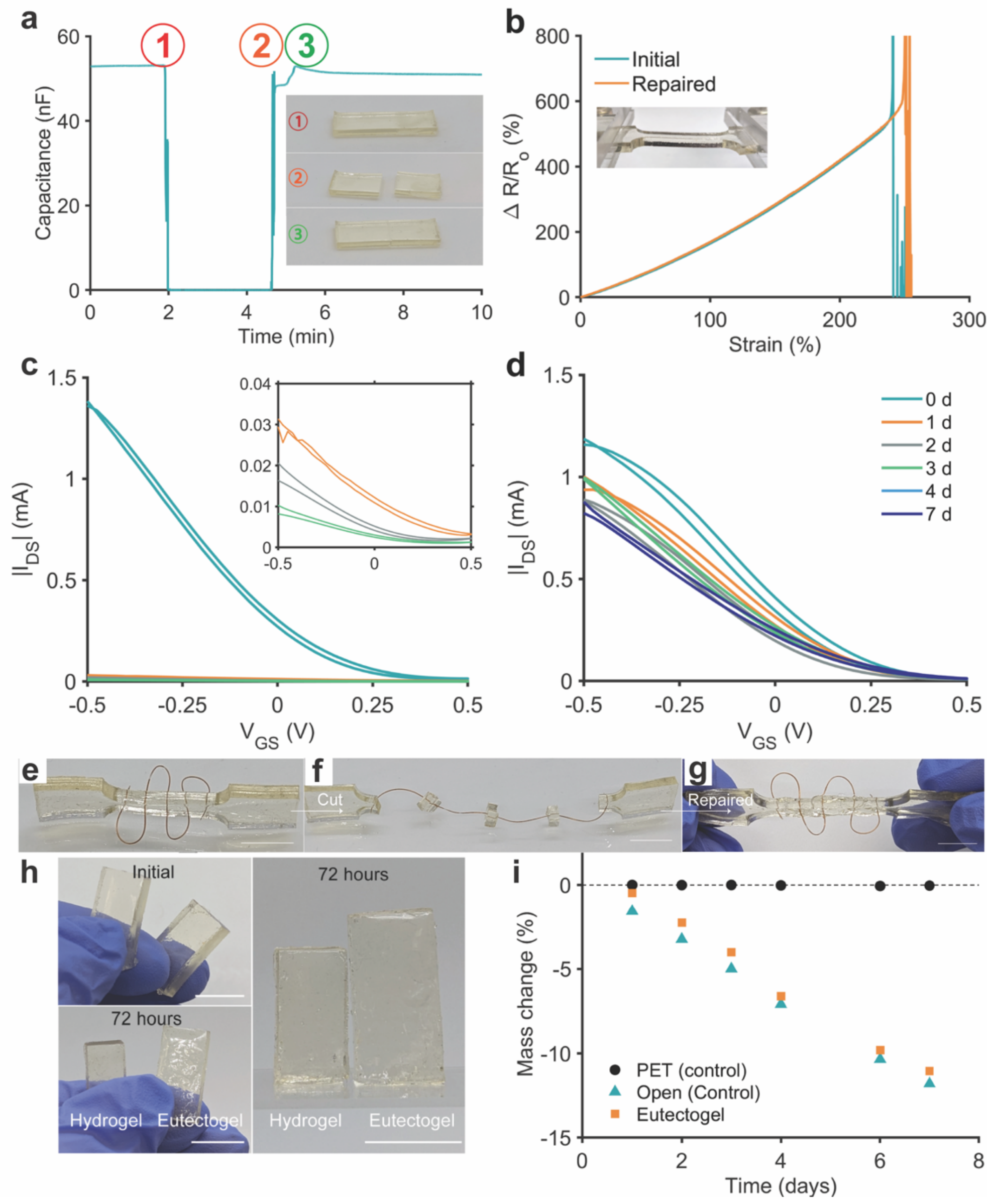

**Figure 5. Attributes of Eutectogels for wearable applications. a.** Restoration of the eutectogel capacitance after a catastrophic break in the gel with photograph inset of a eutectogel at corresponding steps along the graph. **b.** Series resistance of the eutectogel during strain until break. **c.** Transfer curves from a choline chloride hydrogel gated OEGET over time with inset to view results after 1-3 days. **d.** Transfer curves from a eutectogel-gated OEGET over time. **e.** Photograph of a 5-transistor daisy-connected OEGET, which is **f.** catastrophically broken, then **g.** repaired to its initial state via heating of the broken interfaces. Functionality is fully recovered **h.** Photographs of the hydrogel and eutectogel initially and after 3 days. Scale bars are all 1 cm. **i.** Water vapor permeability of a eutectogel.

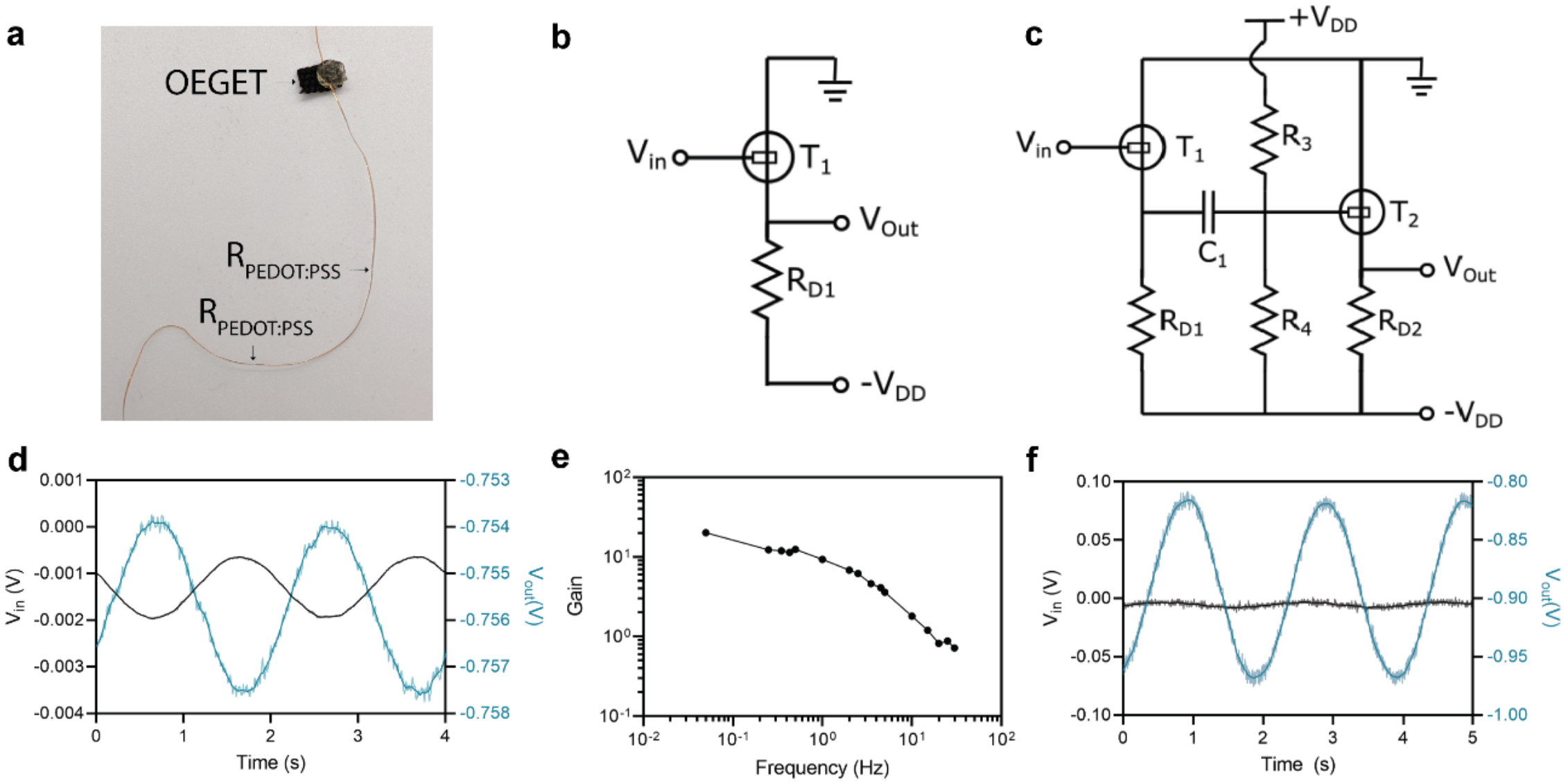

**Figure 6. a.** Photograph of single thread OEGET amplifier. It consists of a transistor and two resistors in series all on a single thread **b.** Electrical schematic of a common source OEGET amplifier. **c.** Electrical schematic of a two stage OEGET cascade. **d.** Amplified signal from the low voltage single thread OEGET amplifier. **e.** Gain versus frequency for a single stage common source OEGET amplifier. **f.** Amplified signal from the two stage OEGET cascade.

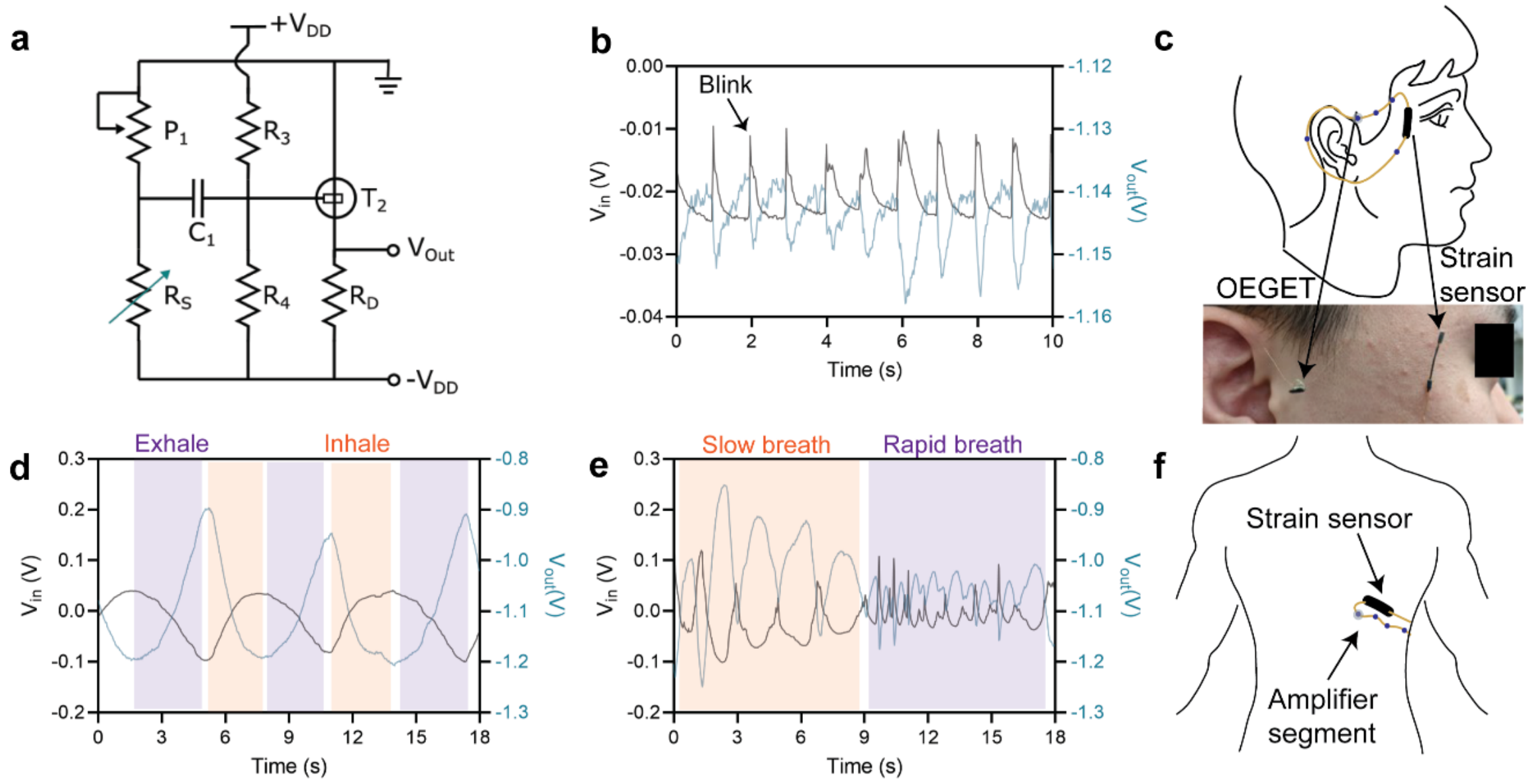

**Figure 7. a.** Electrical schematic of integrated amplified strain sensing circuit. The strain sensor was realized using previously published thread-based strain sensor. This is shown with a resistor $R_s$ in the schematic. **b.** Circuit amplified blinking signal. **c.** Cartoon illustration of sensor placement for monitoring blinking with actual photograph inset. **d.** Circuit amplified inhale and exhale breathing patterns at 20 bpm and **e.** circuit amplified breathing patterns with varied breathing rates. **f.** Cartoon illustration of sensor placement on the diaphragm for monitoring of breathing using the same thread-based strain sensor discussed earlier.

**Supporting Information for**

# Free form three dimensional integrated circuits and wearables on a thread using organic eutectogel gated electrochemical transistors

Rachel E. Owyeung[1,2,3], Wenxin Zeng[2,3], Matthew J. Panzer[1], Sameer Sonkusale[1,2,3*]

[1]Department of Chemical and Biological Engineering, Tufts University, 4 Colby Street, Medford Massachusetts 02155, United States

[2]Department of Electrical and Computer Engineering, Tufts University, 161 College Ave, Medford Massachusetts 02155, United States

[3]Nano Lab, Advanced Technology Laboratory, Tufts University, 200 Boston Ave. Suite 2600, Medford Massachusetts 02155, United States

*Corresponding author: Sameer Sonkusale **Email:** sameer@ece.tufts.edu

**This PDF file includes:**

Supporting text
Figures S1 to S9
Tables S1
SI References

## Supporting Information Text

**Strain sensor fabrication.** The strain sensor was fabricated as reported previously.(1) Briefly, the Gütermann elastic threads (64% Polyester, 36% Polyurethane) were stretched. While under strain, the elastic threads are coated with C-200 Carbon Resistive Ink (Applied Ink Solutions) to aid in the uniformity of coating. Then, the thread is baked in an atmospheric oven at 80°C for 15 minutes. The conductive threads are then coated by EcoFlex to protect the carbon coating.

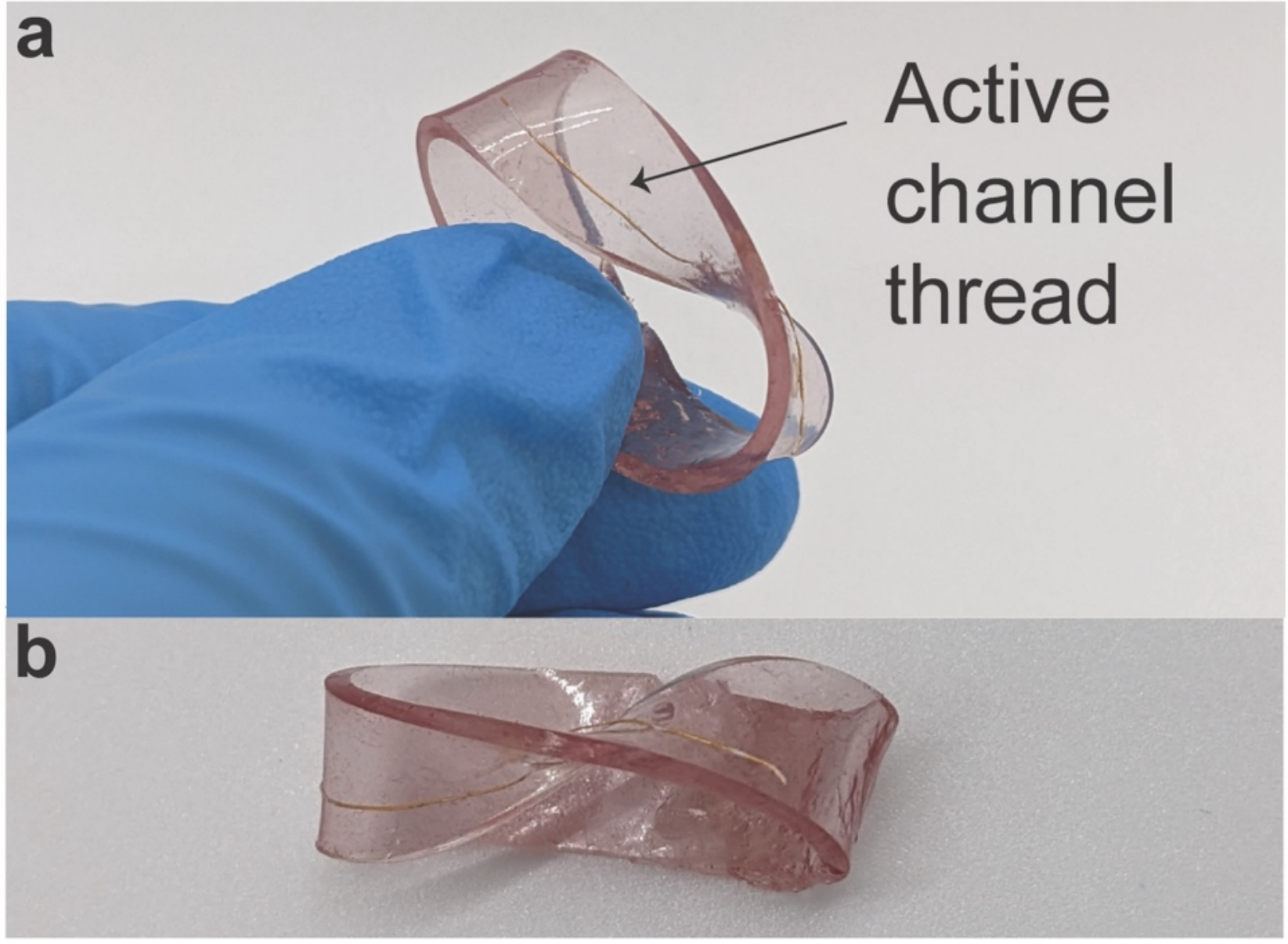

**Figure S1.** Alternate realization of OEGET and OEGET IC showing transistor does not have to confine to two dimensional pattern **a.** and **b.** Example OEGET shaped into a Möbius strip. The eutectogel has been dyed for better visualization.

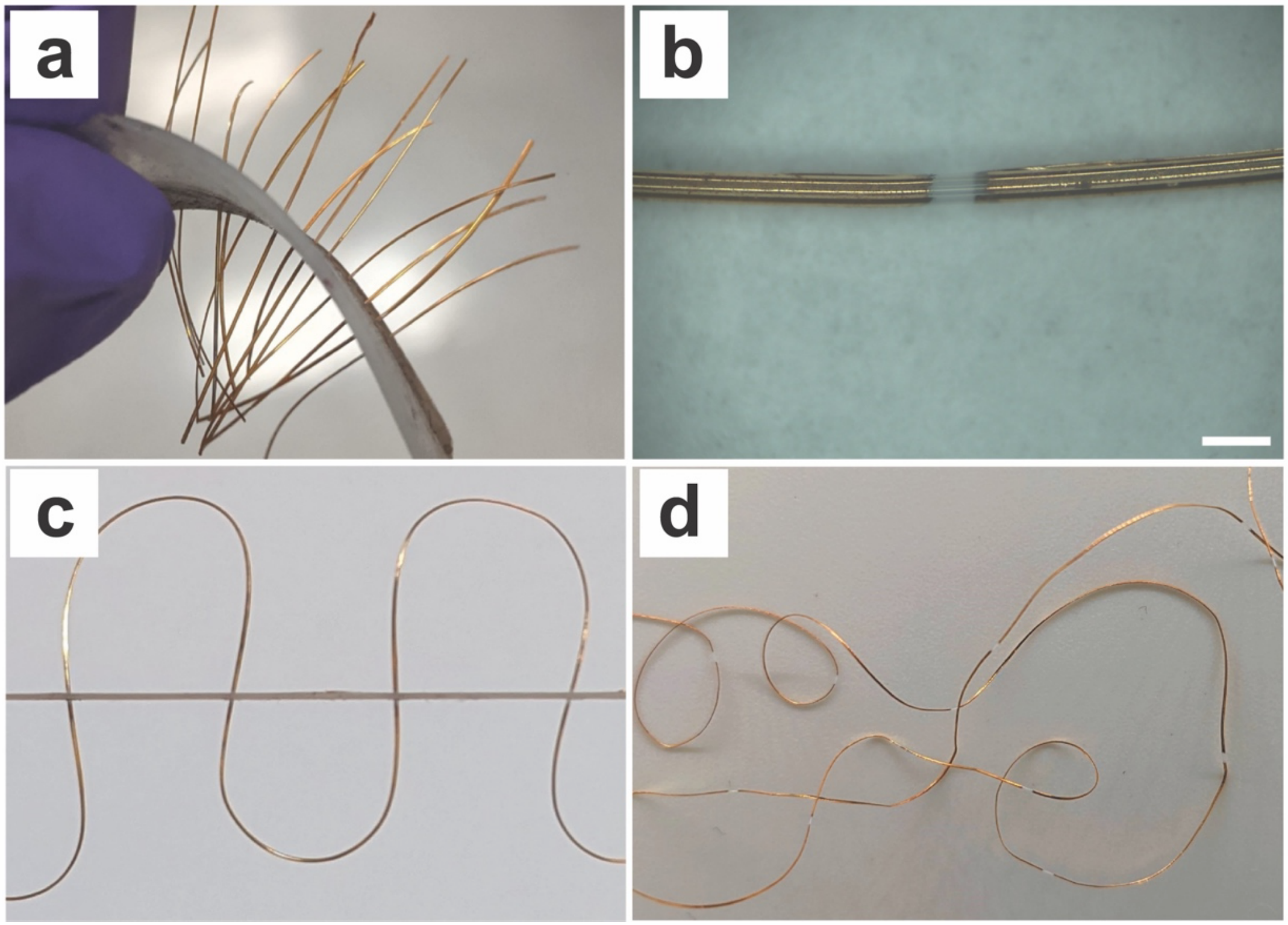

**Figure S2. a.** Fiber/mask with each individual transistor fiber **b.** Resulting single thread showing active channel gap **c.** Side view of fiber mask with daisy connected transistors. Pulling out of the mask results in free form thread with active channel gaps in **d.**

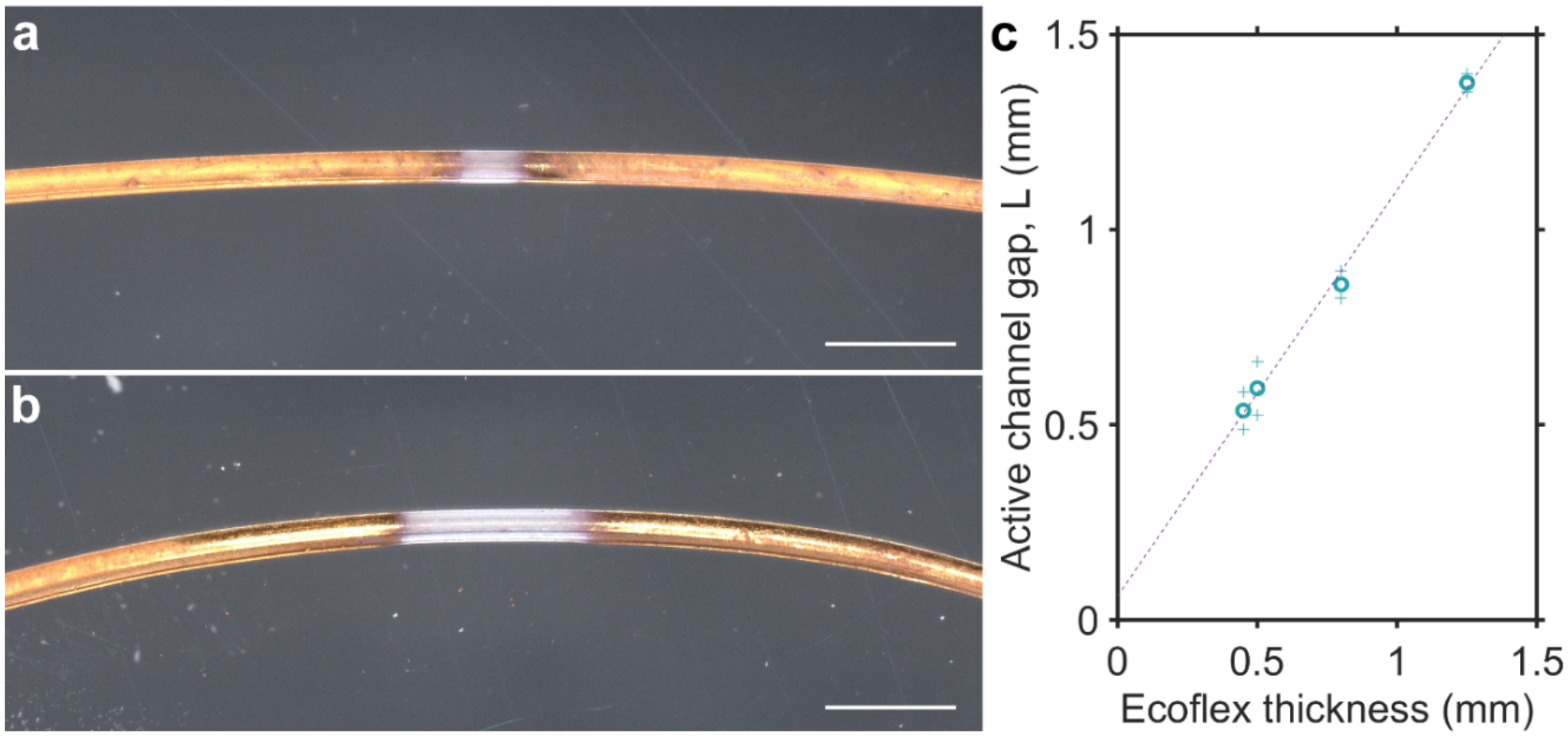

**Fig. S3.** Optical Images of gold coated PCL with Ecoflex mask thickness of a. 0.45 mm and b. 1.25 mm. Scale bar is 1 mm. c. Resulting active channel gap versus the thickness of the Ecoflex mask used.

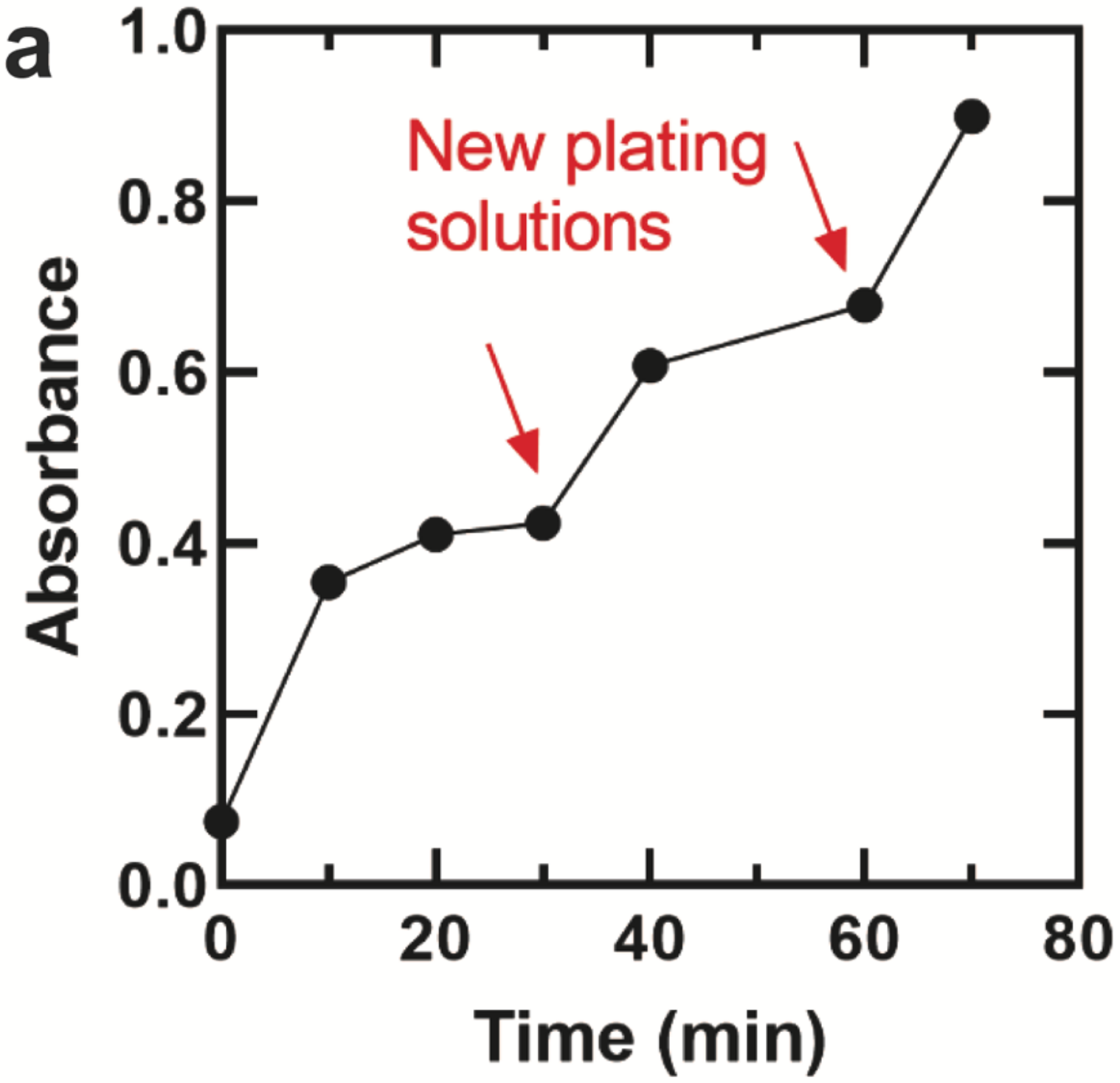

**Fig. S4.** Photographs of gold-coated PCL fibers and UV-Vis absorbance data of the gold electroless deposition process on a glass slide. The gold salt solution was replaced at minute 30 and minute 60, as indicated by the arrows on the graph.

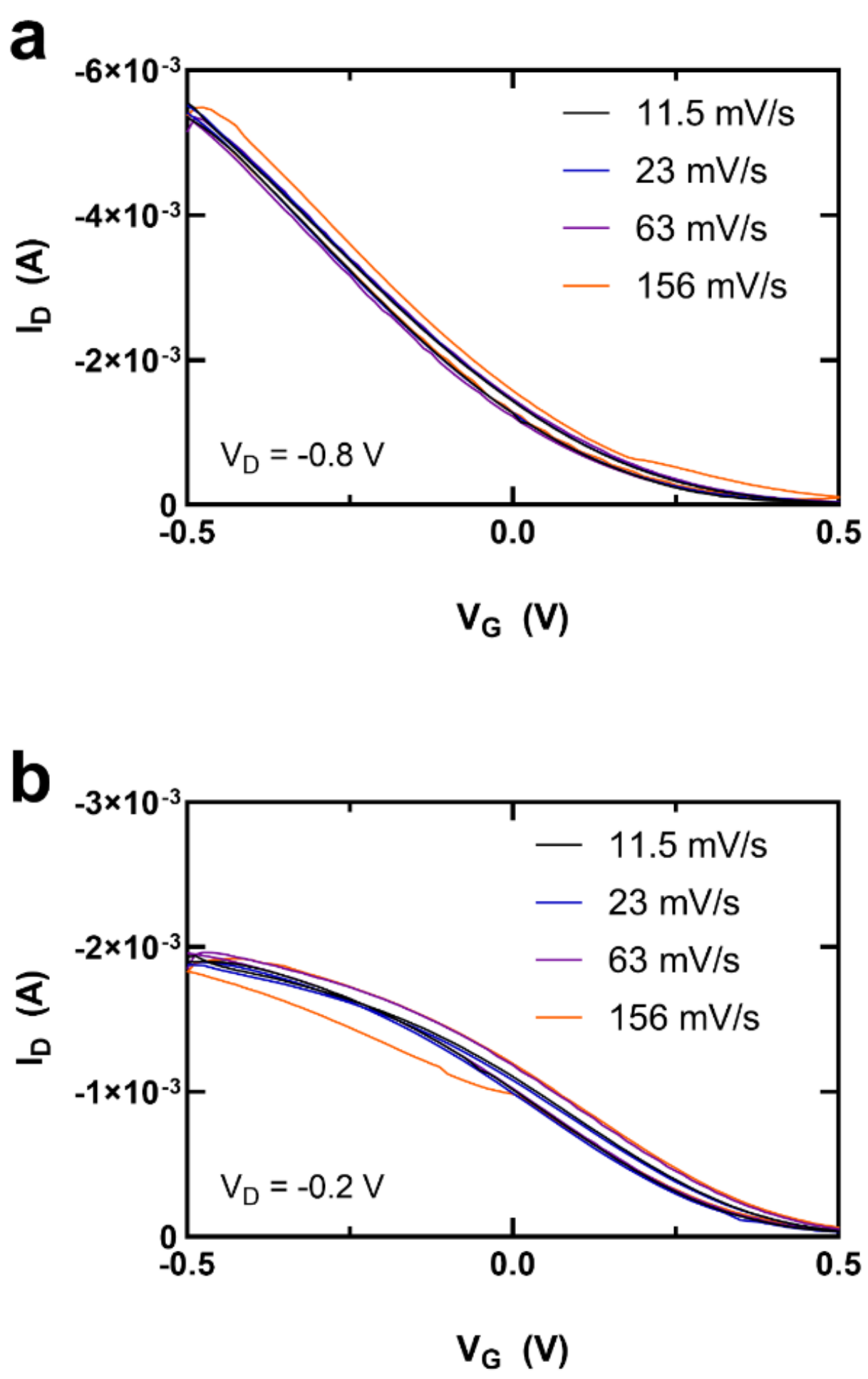

**Figure S5.** Sweep rate dependent hysteresis of eutectogel OECTs in (a) Saturation (VDS = -0.8V) and (b) Linear (VDS = -0.2V) mode

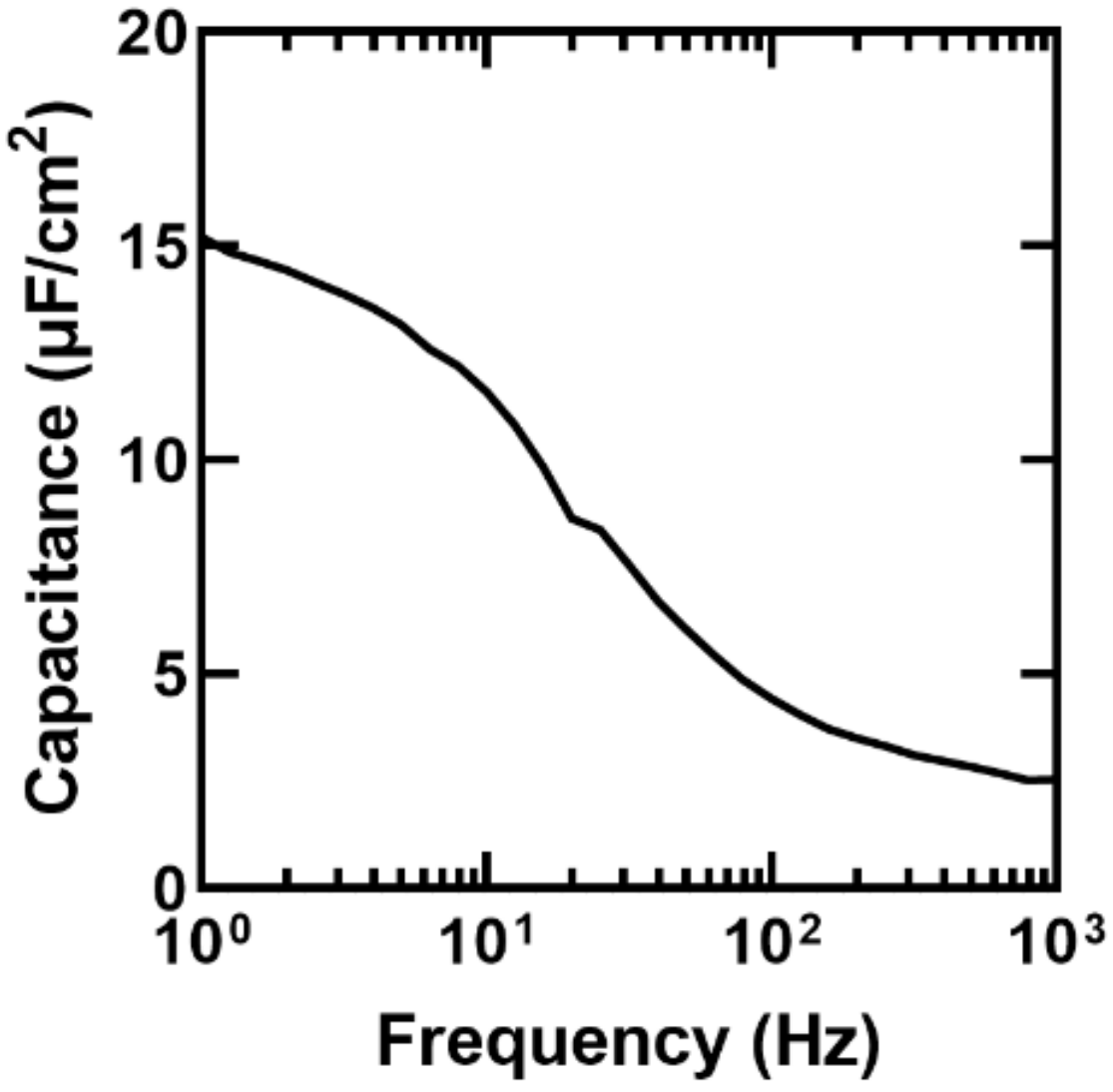

**Figure S6.** Specific Capacitance of the eutectogel as a function of frequency. The gel was sandwiched between two tin-doped indium oxide (ITO) coated glass slides separated by a poly(tetrafluoroethylene) spacer.

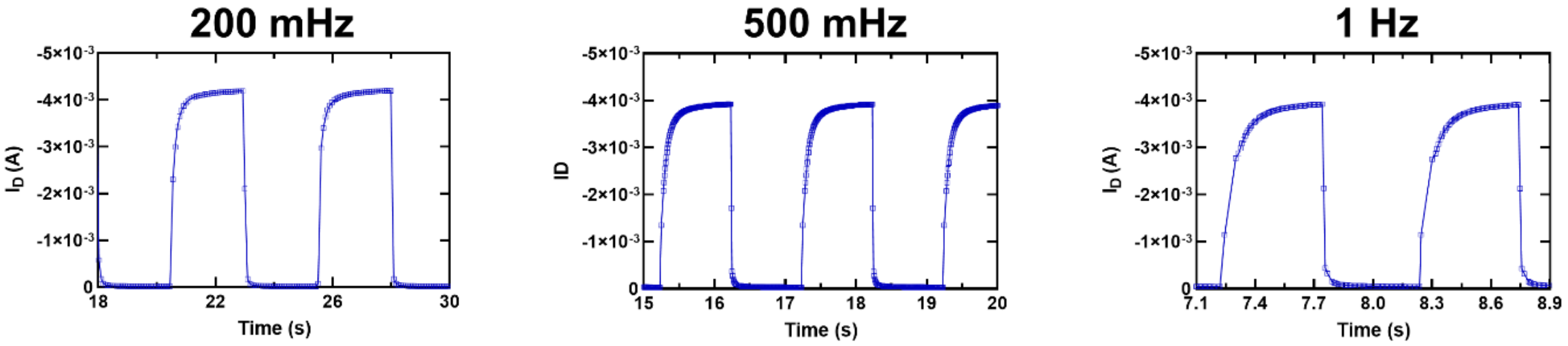

**Figure S7.** Current transient response of an OEGET depending on input voltage frequency

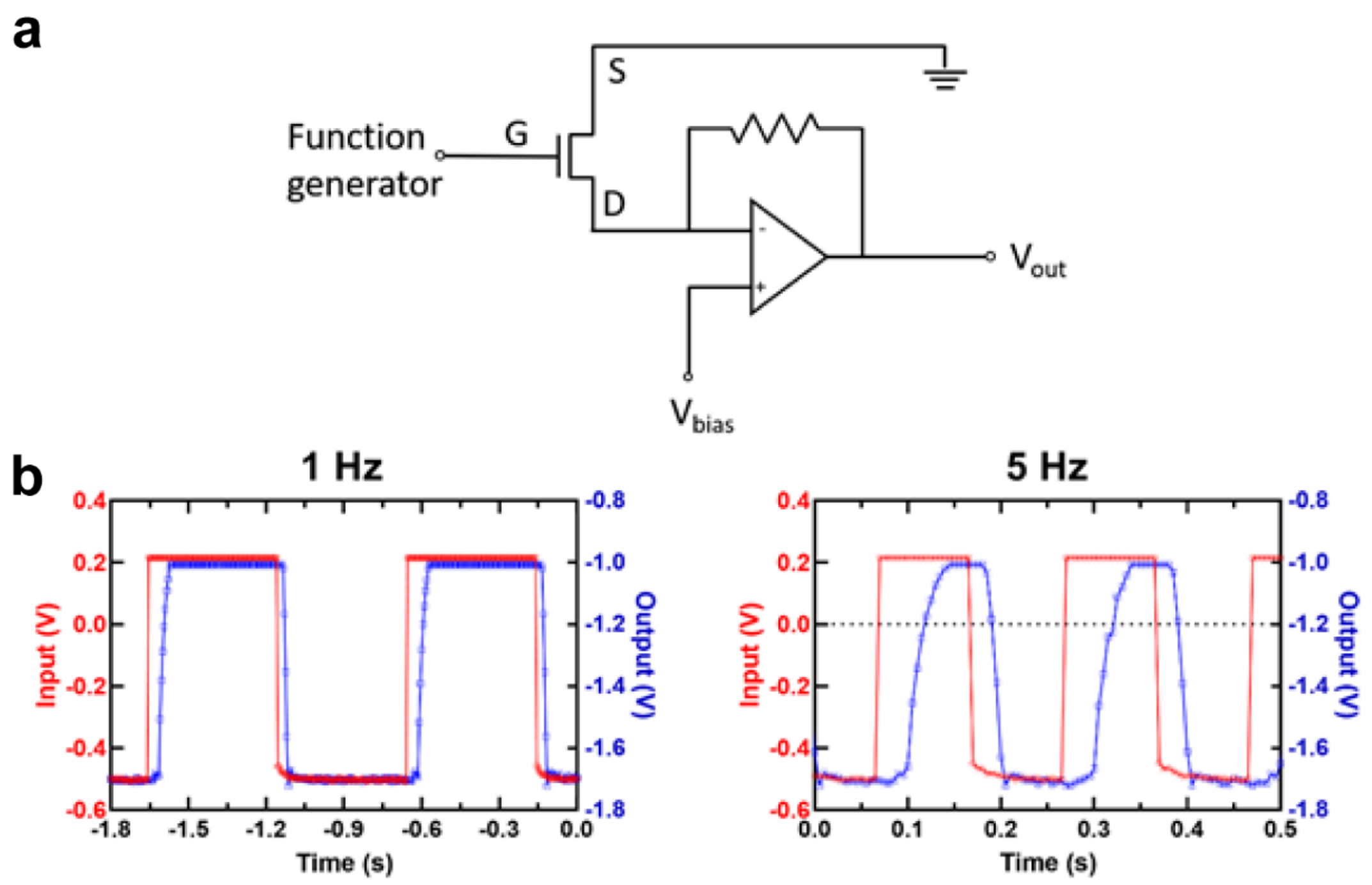

**Figure S8.** Transient response measured via Transimpedance amplifier. a. Transimpedance amplifier schematic. $V_{bias}$ = -0.8V applied via Keithley 2400 source meter. Function generator applied $V_G$ = -0.5 to 0.2V with a square wave. b. Transient response of OEGET with a 1 Hz or 5 Hz square wave input.

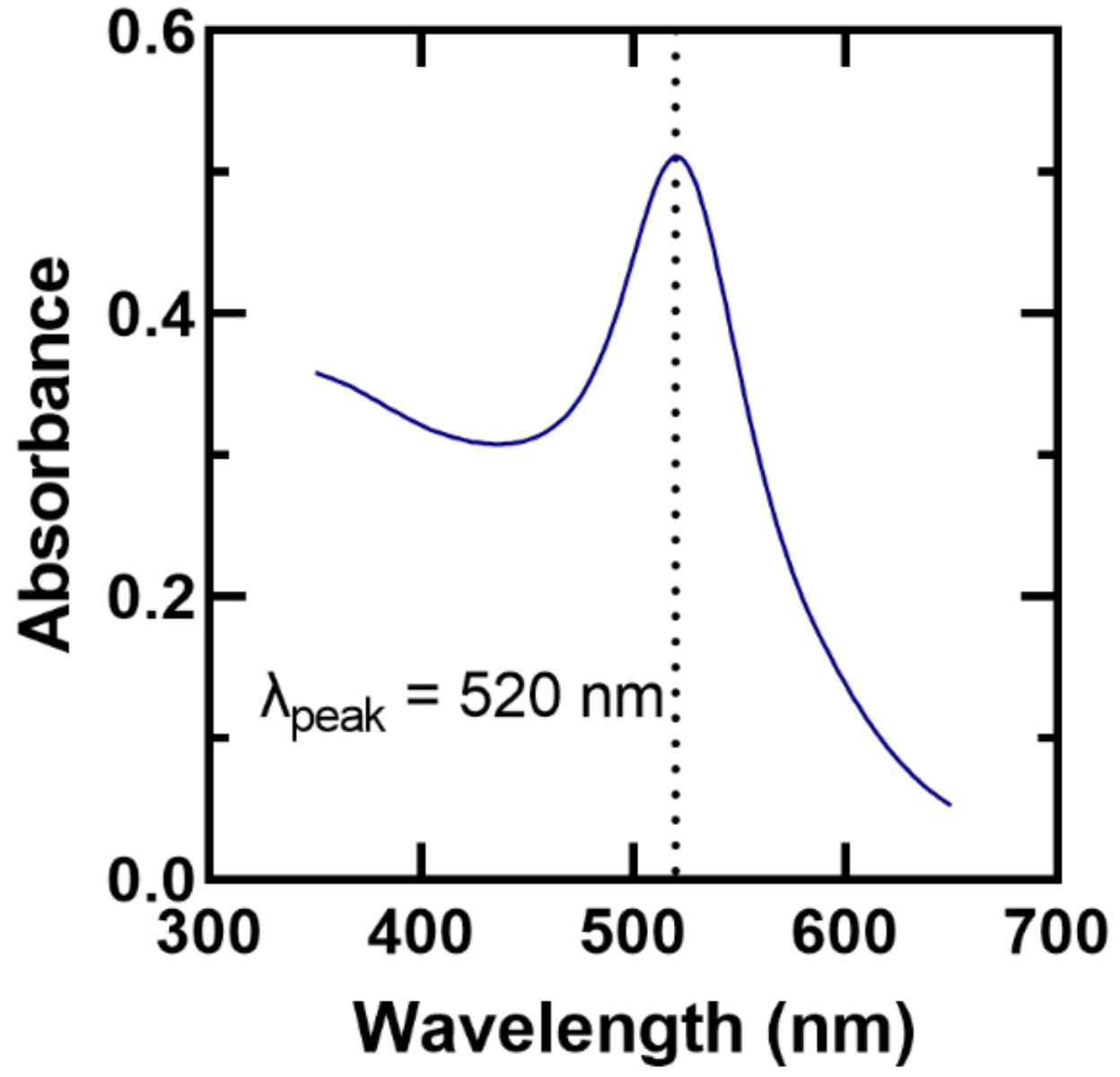

**Figure S9.** Surface plasmon resonance of gold nanoparticle solution.

**Table S1.** Comparison of PEDOT:PSS OECT attributes relevant to wearable electronic circuits. "x" indicates either experiments were not performed, or do not meet the criteria. "-" indicates data not explicitly given.

| Ref. | Electrolyte | Substrate/ geometry | $g_m$ (mS) | ON/ OFF | Wd/L (µm) | Circuits? | Flexibility | Stability | Healing |
|---|---|---|---|---|---|---|---|---|---|
| (1) | NaOH/NaCl/malic acid hydrogel | PET/ planar | 0.27 | $10^2$ | 0.04 | Logic gates, inverter (gain of 1.5 at 0.575 V input voltage) | x | 5 days | x |
| (2) | 0.1M NaCl solution | Glass slides/ planar | - | - | 0.14-1.0 | Amplifier - Gain of >20 V/V for champion device | x | x | x |
| (3) | 0.1M NaCl solution | Glass slide/ planar | 1.5 | $10^5$ | 0.12 | Amplifier - Gain 12 V/V | x | x | x |
| (4) | 0.1M NaCl solution | Polyimide supported by PDMS coated PET/planar | ~6 | $10^5$ | 12.9 | x | Yes | x | x |
| (5) | 0.1 M $NaClO_4$ hydrogel | Polyamide monofilaments/ Cross-fiber | - | $10^3$ | - | Logic gates | x | x | x |
| (6) | 0.1 M NaCl solution | Cotton yarn/ parallel fibers | - | - | - | x | x | x | x |
| (7) | 0.1M NaCl solution | Si wafer (or SEBS for strain test)/planar | 14.2 | $10^3$ | 2 | x | Yes | 1 mo. active channel exposed to ambient air, GOPs crosslinker used | x |
| (8) | 0.1 M NaCl hydrogel | PEDOT:PSS fiber/ cross-fiber | 19 | - | - | Inverter – Gain ~3 V/V | Yes | x | x |
| (9) | triisobutyl(methyl) phosphonium tosylate + $H_2O$ | Glass slide/ planar | - | $10^3$ | 0.05 | x | x | x | x |
| (10) | [EMI][OTF] PVDF-co-HFP ionogel | SEBS/ planar | ~13 | $10^3$ | 0.8 | x | Yes | x | Self-healing |
| (11) | 0.1M NaCl solution | Si wafer or PDMS substrate/planar | 54 | $10^3$ | 10 | x | x | Stored at 70% RH for 68 days, active channel exposed to air, GOPs crosslinker used | Self-healing PEDOT:PSS with triton X-100 |
| **This work** | **ChCl eutectogel** | **Free-form geometry** | **16** | **$10^3$** | **140** | **Single stage - 20 V/V (50 mHz) Two stage - 34 V/V (500 mHz)** | **Yes** | **> 1 mo with electrolyte present continuously, no crosslinker** | **Repair of gel electrolyte** |

**SI References**